\documentclass[10pt]{article}
\usepackage{textcomp}
\usepackage{a4wide}
\usepackage{graphicx}
\usepackage{amsmath}
\usepackage{amsbsy}
\usepackage{epstopdf, epsfig}
\usepackage{amssymb}
\usepackage{longtable}
\usepackage{mathtools}
\usepackage{textcomp, csquotes}
\usepackage{xfrac} 
\usepackage[inline]{enumitem}
\usepackage{afterpage}
\usepackage{natbib}

\usepackage{gensymb}
\usepackage{setspace}
\usepackage{geometry}
\usepackage{lineno}
\usepackage[hidelinks,colorlinks=true,linkcolor=blue]{hyperref}
\hypersetup{citecolor=blue, urlcolor=blue}
\usepackage{ulem}
\usepackage{multirow}
\makeatletter
\def\squiggly{\bgroup \markoverwith{\textcolor{blue}{\lower3.5\p@\hbox{\sixly \char58}}}\ULon}
\makeatother

\title{Practical approaches to study microbially induced calcite precipitation at the field scale}

\author{D. Landa-Marb\'an$^{1}$ \and
  S. Tveit$^{1}$ \and
  K. Kumar$^{2}$ \and
  S.E. Gasda$^{1}$}
\date{}
\begin{document}
\normalem
\pagenumbering{arabic}
\maketitle
\onehalfspace
\noindent ${}^1$ NORCE Norwegian Research Centre AS, Nyg{\aa}rdsgaten 112, 5008 Bergen, Norway.\\[5pt]
${}^2$ Department of Mathematics, Faculty of Mathematics and Natural Sciences, University of Bergen, All\'egaten 41, 5020 Bergen, Norway.\\[5pt]
Corresponding author: David Landa-Marb\'an (E-mail address: dmar@norceresearch.no).
\vspace{1.2cm}
\begin{abstract}
\noindent Microbially induced calcite precipitation (MICP) is a new and sustainable technology which utilizes biochemical processes to create barriers by calcium carbonate cementation; therefore, this technology has a potential to be used for sealing leakage zones in geological formations. The complexity of current MICP models and present computer power limit the size of numerical simulations. We describe a mathematical model for MICP suitable for field-scale studies. The main mechanisms in the conceptual model are as follow: suspended microbes attach themselves to the pore walls to form biofilm, growth solution is added to stimulate the biofilm development, the biofilm uses cementation solution for production of calcite, and the calcite reduces the pore space which in turn decreases the rock permeability. We apply the model to study the MICP technology in two sets of reservoir properties including a well-established field-scale benchmark system for CO$_2$ leakage. A two-phase flow model for CO$_2$ and water is used to assess the leakage prior to and with MICP treatment. Based on the numerical results, this study confirms the potential for this technology to seal leakage paths in reservoir-caprock systems.  
\end{abstract}

\section{Introduction}
\begin{figure}[h!]
\centering
\includegraphics[width=.75\textwidth]{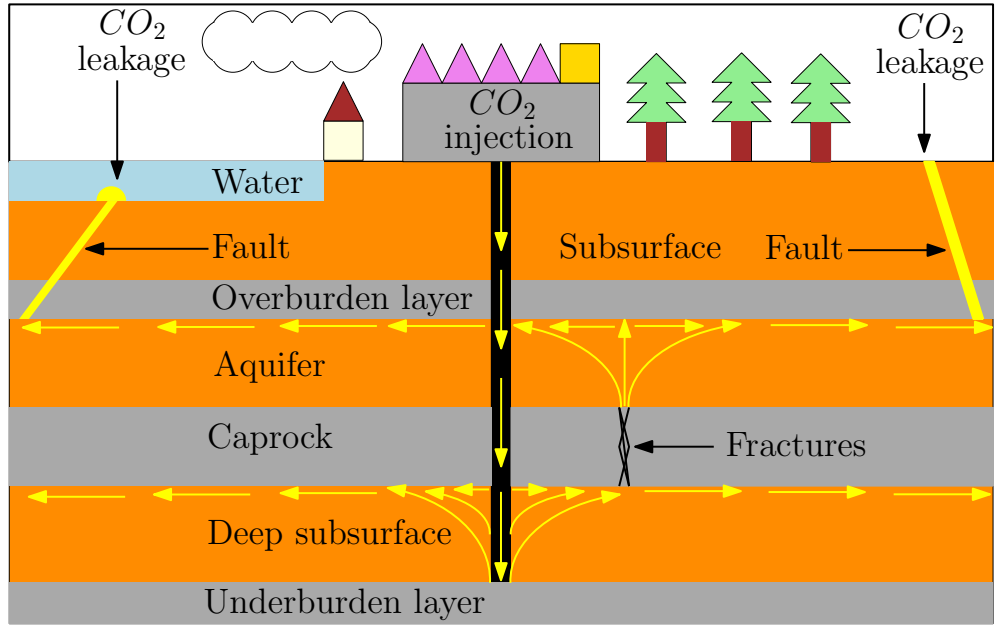}
\caption{Contamination of water and atmosphere by CO$_2$ leakage.}
\label{modelco} 
\end{figure}

Negative emissions technologies and carbon storage must be implemented to avoid dangerous climate changes \citep{Haszeldine:Article:2018,Tong:Article:2019}. Carbon capture and storage (CCS) is one of the promising scalable technologies for storing huge amounts of CO$_2$. Indeed, large amounts of CO$_2$ have already been stored in geological formations on the Norwegian continental shelf, e.g., in the Sleipner field, where more than 16 Mt CO$_2$ has been stored since 1996 \citep{Furre:Article:2017}. Caprocks in reservoirs provide the main trapping mechanism for CO$_2$ sequestration \citep{Bentham:Article:2005}. The existence of faults, fractures, and abandoned wells in the primary sealing caprock of a CO$_2$ storage reservoir can create pathways for CO$_2$ to migrate back to the surface \citep{Fang:Article:2010}. Fig. \ref{modelco} shows a schematic representation of CO$_2$ sequestration, where fractures in the caprock are a risk to leak CO$_2$ back to the atmosphere and to fresh water.

It is therefore necessary to develop methods for mitigating CO$_2$ leakage to ensure its long-term storability. One of the proposed remediation measures to seal leakage zones is the use of microbes to induce precipitation of calcium carbonate \citep{Phillips:Article:2016}. Microbially induced calcite precipitation (MICP) is a new and sustainable technology which utilizes biochemical processes to create barriers by calcium carbonate cementation. The MICP technology involves the injection of diverse components into a reservoir such as microbes, growth solution, and chemicals. As calcite permeability is very low, then the formation of calcite decreases the rock permeability. Thus, MICP technology has a potential to be used for sealing leakage zones in geological formations. These barriers can significantly reduce CO${}_2$ leakage even when the leakage channels are not fully plugged \citep{Li:Article:2019}. Among other applications of MICP besides as a leakage prevention tool in CO$_2$ sequestration are in biomineralized concrete \citep{Lee:Article:2018}, improvement in the stiffness and strength of granular soils \citep{Jalili:Article:2018,Whiffin:Article:2007}, wastewater treatment \citep{Torres-Aravena:Article:2018}, and erosion control \citep{Jiang:Article:2017}.  

MICP as a leakage mitigation technology is intended for use on the field scale, but performing field-scale experiments is expensive. Experiments on microsystems allow us to observe processes in more detail, which leads to improvements in core-scale experiments prior to field applications. We mention some notable works in this direction. \cite{Bai:Article:2017} performed MICP experiments in microfluidic cells to study the distribution of calcite precipitation at the pore scale and observed that calcite precipitation occurs mainly on the bottom of biofilms. Core samples from reservoirs can be used to study changes in permeability due to biofilm growth and calcite precipitation. For example, \cite{Whiffin:Article:2007} conducted a core-scale experiment to evaluate MICP as a soil strengthening process. Since the laboratory experiment was conducted under field conditions and a significant improvement of strength was observed along the column, the authors concluded that MICP can be used for large-scale applications. \cite{Ebigbo:Article:2012} performed core-scale experiments under controlled conditions for studying the effect of calcite precipitation in porous media. The authors tested different injection strategies to obtain a homogeneous distribution of calcite precipitation along sand-packed columns. Their work provides a successful injection strategy for this purpose and experimental data of four columns. \cite{Mitchell:Article:2013} investigated the MICP processes in a core sample inside a high pressure flow reactor including supercritical CO$_2$ to simulate field conditions. Their experimental results show that MICP can be applied in the presence of supercritical CO$_2$. \cite{Gomez:Article:2017} performed experiments in tanks of 1.7 m diameter and with three wells to study the reactive transport of MICP. Their results show that indigenous microorganisms could be stimulated for MICP in field-scale applications. Based on these and more experimental work reported in literature, mathematical models of this technology can be built for further studies. 

Mathematical models of MICP are important as they help to predict the applicability of this technology and to optimize its benefits. \cite{Zhang:Article:2010} introduced a comprehensive pore-scale model for MICP which includes chemistry, mechanics, thermodynamics, fluid, and electrodiffusion transport effects. The authors performed simulations under different conditions of flow rates, concluding that the flow significantly impact the calcite distribution. \cite{Hommel:Article:2015} introduced a core-scale mathematical model for MICP which includes chemistry, mechanics, and fluid transport effects. The authors also calibrated some of the model parameters with experimental data. \cite{Minto:Article:2019} proposed a mathematical model for MICP and performed numerical studies of calcite precipitation around a production well using eight surrounding injection wells. The authors concluded that uniform calcite precipitation could be achieved by splitting the injection into phases, where different number of wells are used in each of the injection phases. Note that \enquote{phases} is used to denote both physical phases and repeatable steps in the injections strategies, and the meaning will be clear from the context.

Despite advances in modeling, simulation of the MICP process at the field-scale is challenging as current mathematical models involve the solution of large systems of highly coupled partial differential equations. In \cite{Cunningham:Article:2019} the authors suggested different approaches to handle this issue such as refinement of the grid locally, multi-scale methods, improving the time stepping, or reducing the coupling of the model equations. \cite{Tveit:Article:2018} proposed a simplified version of the MICP model presented in \cite{Hommel:Article:2015} to perform field-scale simulations. The authors studied two different approaches for inducing calcite precipitation at a given distance of an injection well. Since the complexity of current MICP models and present computer power limit the size of numerical simulations, then simplified models are needed to perform field-scale studies. In \cite{Hommel:Article:2016} the authors discussed a few well-chosen model reductions for the MICP process such as simplification of physics and chemistry, fewer components, and considering a single-phase system. In this work, we build a single-phase field-scale model of MICP technology. This model includes the transport of dissolved components (suspended microbes, growth solution, and cementation solution), biofilm activity (microbial attachment, death, detachment, and growth), and production of calcite which reduce the rock porosity and hence the effective permeability. We use the model to investigate the prevention and sealing of leakage paths located at a certain distance away from the injection well. A simple two-phase flow model for CO$_2$ and water is used to assess the leakage prior to and with MICP treatment.

Our motivation to develop the mathematical model and numerical tools is as follows. We aim to have a model that captures the key processes and quantities involved in the MICP process. At the same time, we aim to have a model which is simple enough so that  computational costs are less. Our main reason for the latter is that the field-scale processes require running the model on a large scale and also require multiple simulations to perform optimization studies. All these imply a heavy computational burden unless we simplify the model. Needless to say, the simplified model should still retain the essence of the processes so that it is useful. Our work is therefore a step in this direction of achieving the twin objectives. 

The paper is structured as follows. In Section 2 we explain in detail the MICP mathematical model, the model parameters, the computer implementation of the model, and the injection strategy. Diverse field-scale numerical experiments to prevent CO$_2$ leakage using MICP are presented in Section 3. A discussion on the numerical results and findings is given in Section 4. Finally, we present the conclusions in Section 5.   
 
\section{MICP model}\label{sec:MICP}
\begin{figure}[h!]
\centering
\includegraphics[width=\textwidth]{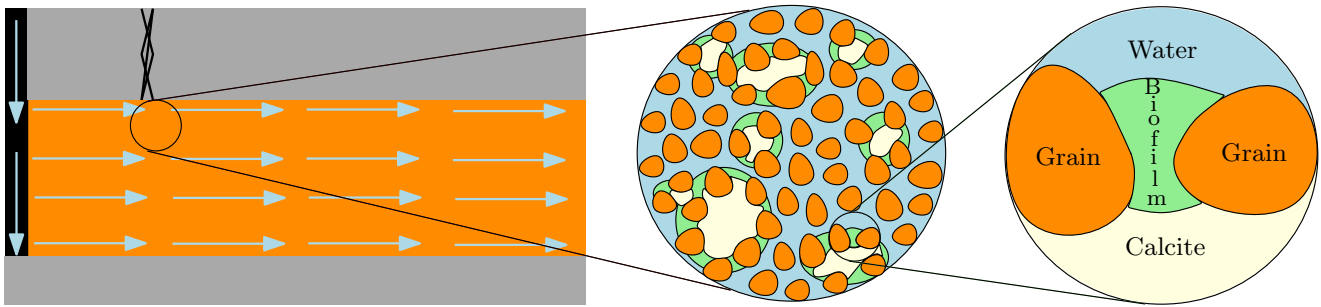}
\caption{Sealing leakage paths with calcite using MICP.}
\label{sealing} 
\end{figure}

In this section we describe the mathematical model for MICP, introducing first concepts and definitions related to this technology. Fig. \ref{sealing} shows a schematic representation of the sealing mechanism using MICP. Here, we observe a fractured zone in the caprock being remediated by calcite. 

MICP can be defined as a bio-geochemical process which results in precipitation of calcite (the low-pressure, hexagonal form of $\text{CaCO}_3$). Calcium carbonate ($\text{CaCO}_3$) is a mineral that naturally precipitates as a result of microbial metabolic activities. Biofilm formation is a process whereby microorganisms attach themselves to a surface and produce an adhesive matrix of extracellular polymeric substances (EPS). In the following we denote growth solution as the mix of components a biofilm needs to develop such as electron acceptors, glucose, nutrients, and substrates. Urea [CO(NH$_2$)$_2$] is a water-soluble compound found in the urine and other bodily fluids of mammals or produced synthetically. Urease is an enzyme catalyzing the hydrolysis of urea to ammonium (NH$_4^+$) and carbonate (CO$_3^{2-}$). \emph{Sporosarcina pasteurii} is a non-pathogenic bacterium commonly used for the MICP process as it shows a high urease enzyme activity \citep{Bhaduri:Article:2016}. Calcium ions (Ca$^{2+}$) are important mediators of a wide range of cellular activities, contributing to the biochemistry of microorganisms. Lastly, we denote cementation solution as the injected chemicals, urea, and calcium (e.g., in form of calcium chloride) needed to facilitate the MICP process. With the main concepts introduced, we proceed to describe the conceptual and mathematical model of MICP used in this paper.

\subsection{Conceptual model}
We consider a constant-temperature reservoir saturated with water, where calcite and biofilm only occur on the rock walls, i.e., in the space domain there are one liquid phase (water) and three solid phases (biofilm, calcite, and rock matrix). The microbial medium, growth components, and cementation solution are dissolved in water prior to injection and they are transported only in the water phase by advection and dispersion. The biofilm and calcite are assumed to be impermeable and incompressible. The governing processes in the biofilm are growth, death, attachment, and detachment. We consider the limiting factor in the growth solution to be oxygen (electron acceptor). This assumption can be justified since oxygen has a limited solubility in water \citep{Raimbault:Article:1998}, while the other components can be injected at high concentrations with the growth solution.

The most studied MICP process is urea hydrolysis (ureolysis) via the enzyme urease produced by special microbes, in a calcium-rich environment \citep{Rong:Article:2012,Whiffin:Article:2007}:
\begin{gather*}\label{chemistry}
\begin{split} 
&\text{CO(NH}_2)_2+2\text{H}_2\text{O}\xrightarrow{\text{microbes}} 2\text{NH}_4^++\text{CO}_3^{2-},\hspace{.5cm}\text{ureolysis},\\
&\text{Ca}^{2+}+\text{CO}_3^{2-}\longrightarrow\text{CaCO}_3\downarrow,\hspace{2.9cm}\text{calcite precipitation}.
\end{split}
\end{gather*}

In general, CaCO$_3$ precipitation is governed by four main factors \citep{Hammes:Article:2002}: calcium concentration, carbonate concentration, pH, and availability of nucleation sites. \cite{Lauchnor:Article:2015} performed experiments on \emph{S. pasteurii}, showing that urea and microbial concentration have a more significant impact on the ureolysis rate than pH variations. In addition, we consider that the amount of urease is only related to the amount of biofilm, neglecting the suspended microbes in the liquid phase as their contribution is minor \citep{Ebigbo:Article:2012}. Assuming enough calcium concentration in the water, we model the calcite formation as a function only dependent on urea and biofilm. This assumption can be justified since calcium can be injected together with urea in the cementation solution, and would thus distribute in a similar manner, ensuring that both concentrations are high in the location where calcite precipitation is aimed.

To summarize, the system of interest consists of a 3D reservoir (porous medium), one source (injection well), one fluid phase (water), two solid phases (biofilm and calcite), and three injected solutions (microbial, growth, and cementation solutions). The rate-limiting components in the three injected solutions are suspended microbes, oxygen, and urea respectively.

\subsection{Mathematical model}
We build a mathematical model based on the assumptions laid out in the conceptual model. We adopt a continuum approach, where the processes in the system are described by conservation laws and coupling relationships. We use the subscripts $\lbrace b, c, m, o, u, w\rbrace$ to refer to biofilm, calcite, suspended microbes, oxygen, urea, and water respectively. We emphasize that while the injected solutions are composed of various components (e.g., oxygen, glucose, nutrients, substrates, calcium chloride, pH, and urea), in the mathematical model the rate-limiting components in the growth and cementation solution are oxygen and urea respectively.

\subsubsection{Flow equations} The mass conservation and Darcy's law equations for the water phase are given by:
\begin{equation}\label{tpf}
\frac{\partial \phi }{\partial t}+\nabla\cdot\pmb{v}_w=q_w,\quad \pmb{v}_w=-\frac{\mathbb{K}}{\mu_w}(\nabla p_w-\rho_w\pmb{g}),
\end{equation}
where $\phi $ is the rock porosity, $p_w$ the reservoir pressure, $\pmb{v}_w$ the discharge per unit area, $\rho_w$ the fluid density, $\mathbb{K}$ the absolute permeability, $\pmb{g}$ the gravity, $\mu_w$ the water viscosity, and $q_w$ the source/sink term.

\subsubsection{Leakage paths} We adopt a common approach found in \cite{Class:Article:2009}, where the leakage paths in the caprock are modeled as a porous medium with higher permeability than the formation. An advantage of this approach is that the model equations do not need further modification for implementation while a drawback is that it requires a fine grid to represent explicitly the leakage paths. This may be contrasted with the widely-used approach of discrete fracture networks (DFN) where one uses a mixed dimensional setting and represents the fractures as a $n-1$ dimensional objects embedded in a $n$ dimensional porous geometry. We refer to \cite{Berre:Article:2019} for a recent review of different conceptual models for fracture and \cite{Kumar:Article:2020} and \cite{Martin:Article:2005} for a formal derivation of some of these models. We also mention that our model here can be easily adapted for different conceptual models including fractures being modeled as DFNs.

\subsubsection{Injection well} In this work we consider only one injection well, where the injection rate $Q_w$ is given as follows \citep{Lie:Book:2019,Peaceman:Article:1978}:
\begin{equation}
Q_w=\frac{2\pi \Xi K}{\mu_w \ln(r_e/r_I)}\left[ p_w-p_I-\rho_w(z-z_{bh})g\right ].
\end{equation}

Here, $p_I$ is the pressure inside the wellbore, $\Xi$ is the length of the grid block in the major direction of the wellbore, $r_I$ the well radius, $z_{bh}$ a reference depth, $K$ is the permeability in the direction of the injection, and $r_e$ the radius at which the steady-state pressure for the well equals the numerically computed pressure for the well block. 

\subsubsection{Transport equations} To describe the transport of suspended microbes, oxygen, and urea, we consider the following advection-dispersion-reaction transport equations:
\begin{equation}
\frac{\partial (c_\xi\phi )}{\partial t}+\nabla\cdot \pmb{J}_\xi=c_\xi q_w+R_\xi,\quad \pmb{J}_\xi=-\phi  \mathbb{D}_\xi\nabla c_\xi+c_\xi\pmb{v}_w,\quad \xi\in\lbrace m,o,u\rbrace.
\end{equation}

Here, $c_\xi$ is the mass concentration of component $\xi$ in water, $\pmb{J}_\xi$ the flux of $\xi$, $R_\xi$ the reaction term of $\xi$, and $\mathbb{D}_\xi$ the dispersion tensor. Here, we assume that the aqueous phase density does not depend on the component concentrations.

\subsubsection{Dispersion effects} When the components are transported throughout the reservoir, two different mechanisms affect their movement: mechanical dispersion and molecular diffusion. The former is an effect arising out of mixing due to flow and heterogeneities while the latter accounts for  movement of the components from a region of higher to lower concentration. We adopt the following model for the dispersion of components \citep{Bear:Book:1972}
\begin{equation}
\mathbb{D}_\xi=\alpha_T||\pmb{v}||\mathbb{I}+(\alpha_L-\alpha_T)\frac{\pmb{v}\otimes\pmb{v}}{||\pmb{v}||}+D_\xi\mathbb{I},\quad \xi\in\lbrace m,o,u\rbrace,
\end{equation}
where $\alpha_L$ and $\alpha_T$ are the longitudinal and transverse dispersion coefficients, $\pmb{v}=\pmb{v}_w/\phi $ is the effective velocity of the aqueous phase, and $D_\xi$ the effective diffusion coefficient of component $\xi$.

\subsubsection{Solid-phase equations} As previously mentioned, we consider biofilm formation and calcite precipitation fixed in space (at the pore scale, it represents the biofilm and calcite precipitate at the rock surface). Thus, the following mass balance equations describe the evolution of biofilm and precipitation of calcite 
\begin{equation}
\frac{\partial(\rho_\chi\phi_\chi)}{\partial t}=R_\chi,\quad \chi\in\lbrace b,c\rbrace,
\end{equation}
where $\rho_\chi$ are densities and $R_\chi$ reaction terms which are being described later in this section. 

\subsubsection{Suspended microbes} Two opposing processes determine the evolution of suspended microbes: growth and loss. The growth term comprises of two contributions. First, the consumption of oxygen by the microbes lead to its growth. This is modeled by a Monod equation $c_m\phi Y\mu c_o/(k_o+c_o)$ where $\mu$ is the maximum specific growth rate, $k_o$ the half-velocity coefficient for the oxygen, and $Y$ the growth yield coefficient. Second, its growth taking place via detachment or erosion of biofilm due to flow. Microbes detach from the biofilm back to the water phase due to shear forces on the interface by the water flow. The erosion is modeled by $\phi_b\rho_bk_{str}\phi||\nabla p_w-\rho_w\pmb{g}||^{0.58}$ where $k_{str}$ is the detachment rate \citep{Rittmann:Article:1982}. The loss term also has two contributions. First, the death of the suspended microbes as a result of aging, which is modeled by a linear death rate $-c_m\phi k_d$ where $k_d$ is the microbial death coefficient. Second, the suspended microbes attach themselves to the pore wall and biofilm. This is modeled by a linear attachment rate $-c_m\phi k_a$ where $k_a$ is the microbial attachment coefficient. In sum, the rate for the suspended microbes $R_m$ is given by
\begin{equation}
R_m=c_m\phi \left(Y \mu \frac{c_o}{k_o+c_o}-k_{d}-k_a\right )+\phi_b\rho_bk_{str}\phi||\nabla p_w-\rho_w\pmb{g}||^{0.58}.
\end{equation}

\subsubsection{Oxygen utilization} The oxygen utilization rate $R_o$ is expressed as \citep{Ebigbo:Article:2012}:
\begin{gather}\label{rates}
\begin{split}
R_{o}=-(c_m\phi+\rho_b\phi_b)F\mu \frac{c_o}{k_o+c_o},
\end{split}
\end{gather} 
where $F$ is the mass ratio of oxygen consumed to substrate used for growth.

\subsubsection{Urea utilization} The urea conversion rate $R_u$ is modeled by the Monod equation \citep{Hommel:Article:2015,Lauchnor:Article:2015}
\begin{equation}
R_u=-\rho_b\phi_b\mu_u\frac{c_u}{k_u+c_u},
\end{equation}
where $\mu_u$ is the maximum rate of urea utilization and $k_u$ is the half-velocity coefficient for urea. This model for ureolysis was introduced in \cite{Hommel:Article:2015} based on the work by \cite{Lauchnor:Article:2015}, where $\mu_u$ is split into maximum activity of urease ($k_{\text{urease}}$) and mass ratio of urease to biofilm ($k_{ub}$), i.e., $\mu_u=k_{\text{urease}}k_{ub}$.

\subsubsection{Calcite precipitation} The calcite precipitation is the result of a complex geochemical process. In \cite{Quin:Article:2016} the authors have observed that  in a calcium-rich environment the calcite precipitation rate is limited by the slower ureolysis rate; thus, an approximation of the calcite precipitation rate can be given by the negative value of the urea utilization rate (i.e., $R_c=-R_u$). This simplification on the chemistry process has been compared to experimental data, resulting in a relatively low error in comparison to computing all intermediate reactions \citep{Hommel:Article:2016}. Since the molar mass of urea is different from calcite, we add a yield coefficient $Y_{uc}$ (units of produced calcite over units of urea utilization) to account for this in the mathematical model. Then, we write the calcite precipitation rate as
\begin{equation}\label{calcites}
R_c=\rho_b\phi_bY_{uc}\mu_u\frac{c_u}{k_u+c_u}.
\end{equation}

We note that $ R_c $ only depends on the amount of biofilm and urea, which significantly reduces the computational cost compared to more complex formulations [e.g., \cite{Ebigbo:Article:2012}, \cite{Hommel:Article:2015}, and \cite{Minto:Article:2019}].

\subsubsection{Biofilm processes} As in the case of suspended microbes above, the biofilm development is determined by the net of its growth and loss. Consumption of oxygen by the biofilm lead to its growth. This is modeled by the Monod equation $\rho_b\phi_b Y \mu c_o/(k_o+c_o)$. The microbes in the biofilm die as a result of aging and being encapsulated by the produced calcite \citep{DeMuynck:Article:2010}. The former is modeled by a linear death rate $-\rho_b\phi_bk_d$ while the latter by $-\rho_b\phi_bR_c/[\rho_c(\phi_0-\phi_c)]$ \citep{Ebigbo:Article:2012}. As described previously, the microbial attachment leading to its growth is modeled by $c_m\phi k_a
$ while the erosion leading to its loss is expressed by $-\phi_b\rho_bk_{str}\phi||\nabla p_w-\rho_w\pmb{g}||^{0.58}$. In sum, the rate for the evolution of the biofilm is given by
\begin{equation}
R_b=\rho_b\phi_b\left[Y \mu \frac{c_o}{k_o+c_o}-k_d-\frac{R_c}{\rho_c(\phi_0-\phi_c)}-k_{str}\phi||\nabla p_w-\rho_w\pmb{g}||^{0.58}\right]+c_m\phi k_a.
\end{equation}

\subsubsection{Porosity reduction} The void space in the porous medium change in time as a function of the biofilm and calcite volume fractions $\phi_b$ and $\phi_c$ respectively. Using the definitions of $\phi_b$ and $\phi_c$, we have the following equality
\begin{equation}
\phi =\phi_0-\phi_b-\phi_c.
\end{equation}

\subsubsection{Permeability modification} Porosity-permeability relationships are used frequently in mathematical modeling to account for permeability reduction as a result of biofilm and calcite growth. Diverse porosity-permeability relationships have been proposed for the last decades. These relationships can also include the permeability of biofilm and be derived as a result of upscaling pore-scale models \citep{Landa:Article:2020,vanNoorden:Article:2010}. In this paper, we follow \cite{Thullner:Article:2002} and use a porosity-permeability relationship where significant reduction in CO$_2$ leakage can be achieved even when the leakage paths are not fully plugged,
\begin{equation}\label{Kp}
\mathbb{K}=\begin{cases}
\left[\mathbb{K}_0\bigg(\frac{\phi -\phi_\text{crit}}{\phi_0-\phi_\text{crit}}\bigg)^\eta+K_\text{min}\right]\frac{\mathbb{K}_0}{\mathbb{K}_0+K_\text{min}},&\phi_{\text{crit}}<\phi\\
K_\text{min}, &\phi\leq \phi_{\text{crit}}.
\end{cases}
\end{equation}

Here, $\mathbb{K}_0$ is the initial rock permeability, $\phi_\text{crit}$ is the critical porosity when the permeability becomes a minimum value $K_\text{min}$, and $\eta$ is a fitting factor. 

\subsubsection{Remarks on the MICP model} The development of the present mathematical model is inspired by previous works on the MICP technology \citep{Cunningham:Article:2019,Ebigbo:Article:2012,Hommel:Article:2015,Lauchnor:Article:2015,Quin:Article:2016}. One of the most complete models for the MICP technology is presented in \cite{Hommel:Article:2015}. This MICP model includes detailed chemistry reactions, mechanics, and fluid transport effects. Given the complexity of the model and the current computing power, solving simultaneously all equations would limit the size of the problem. Hence we build a simpler mathematical model so that the computational costs are less. We summarize the main assumptions that we have adopted to build the simplified MICP model: only one fluid phase (water) and three solid phases (biofilm, calcite, and rock matrix) are presented, there are only three rate-limiting components (suspended microbes, oxygen, and urea) dissolved in the fluid phase, the amount of urease is only related to the amount of biofilm, and the calcite formation only depends on urea and biofilm. The mathematical model is given by Eqs. (\ref{tpf}-\ref{Kp}). This model consists of six mass balance equations and six cross coupling constitutive relationships.

\subsection{Implementation}
The EOR module in the MATLAB$^\text{\textregistered}$ reservoir simulation tool (MRST), a free open-source software for reservoir modeling and simulation, is modified to implement the MICP mathematical model \citep{Bao:Article:2017,Lie:Book:2019}. Specifically, the polymer example (black-oil model + one transport equation) is modified (single-phase flow + three transport equations + two mass balance equations) to solve the MICP mathematical model. A comprehensive discussion of the solution of the polymer model can be found in \cite{Bao:Article:2017}. The MICP mathematical model is solved on domains with cell-centered grids. Two-point flux approximation (TPFA) and backward Euler (BE) are used for the space and time discretization respectively. The resulting system of equations is linearized using the Newton-Raphson method. In contrast to the polymer model, we implement dispersion of the transported components, permeability changes due to calcite and biofilm formation, and biofilm detachment due to shear forces. The spatial discretization is performed using internal functions in MRST and the external mesh generator DistMesh \citep{Persson:Article:2004}. The MICP processes can be simulated over time and the simulator stops when full-plugging of at least one cell is reached (i.e., $\phi=0$). The links to download the corresponding code can be found above the references at the end of the manuscript.

\subsection{Model parameters}
Mathematical models require the numerical values of coefficients in the equations to be solved. These model parameters are system-dependent and their values are estimated by different means, e.g., direct measurements of the system and experimental data. Experiments under controlled input quantities aim to provide a better estimation of these parameters. For example, in \cite{Landa:Article:2019} the detachment rate for the bacterium \emph{Thalassospira} strain A216101 was estimated after performing measurements of the biofilm development under different flow rates. The MICP mathematical model consists of 21 model parameters whose value may depend on the species of bacteria, temperature in the system, rock type, etc. In this work, we use model parameters reported in the literature. 

Table \ref{tab:allparameterss} summarizes the model parameters for the numerical simulations. We comment on the maximum rate of urea utilization $\mu_u$, yield coefficient $Y_{uc}$, and minimum permeability $K_\text{min}$. \cite{Lauchnor:Article:2015} estimated values for the kinetics of ureolysis by \emph{S. pasteurii}. We consider a value of $\mu_u=1.61\times 10^{-2}$ s$^{-1}$ [here we use the value of mass ratio of urease to biofilm of $3.81\times 10^{-4}$ and 0.06 kg/mol for urea multiplied by 706.7 mol/(kg s) \citep{Lauchnor:Article:2015}]. The molar mass ratio of calcite (0.1 kg/mol) to urea (0.06 kg/mol) gives a value of 1.67 for the yield coefficient $Y_{uc}$. The value of $K_\text{min}$ is set to $10^{-20}$ m$^2$ which is of the order of magnitude of permeability in a caprock to retain fluids for \mbox{CCS \citep{Schlumberger:Definition:2020}.}\\

\begin{table}[h!]
\centering
\caption{Table of model parameters for the numerical studies.}
\begin{tabular}{ l l l l l}		
\hline
Parameter & Symbol & Value & Unit & Reference\\
\hline
Density (biofilm) &$\rho_b$ & $35$ & $\sfrac{\text{kg}}{\text{m}^3}$&\cite{Hommel:Article:2015}\\
Density (calcite)& $\rho_c$	& $2710$ & $\sfrac{\text{kg}}{\text{m}^3}$&Standard\\
Density (water) &$\rho_w$	& 	$1045$ & $\sfrac{\text{kg}}{\text{m}^3}$&\cite{Ebigbo:Article:2007}\\
Detachment rate & $k_{str}$ & 2.6$\times10^{-10}$ & $\sfrac{\text{m}}{\text{(Pa s)}}$ & \cite{Landa:Article:2019}\\
Critical porosity & $\phi_\text{crit}$ & 0.1& [-]& \cite{Hommel:Article:2013}\\
Diffusion coefficient (suspended microbes)& $D_m$	& $2.1\times 10^{-9}$ & $\sfrac{\text{m}^2}{\text{s}}$ &\cite{Kim:Article:1996}\\
Diffusion coefficient (oxygen)& $D_o$	& $2.32\times 10^{-9}$ & $\sfrac{\text{m}^2}{\text{s}}$ &\cite{Chen:Article:2013}\\
Diffusion coefficient (urea)& $D_u$	& $1.38\times 10^{-9}$ & $\sfrac{\text{m}^2}{\text{s}}$ &\cite{Nanne:Article:2010}\\
Dispersion coefficient (longitudinal)& $\alpha_L$ & 	$10^{-3}$ & m &\cite{Benekos:Article:2006}\\
Dispersion coefficient (transverse) & $\alpha_T$ & 	$4\times10^{-4}$ & m &\cite{Benekos:Article:2006}\\
Fitting factor & $\eta$ & 3& [-]& \cite{Hommel:Article:2013}\\
Half-velocity coefficient (oxygen)&$k_o$	& 	$2\times 10^{-5}$ & $\sfrac{\text{kg}}{\text{m}^3}$&\cite{Hao:Article:1983}\\
Half-velocity coefficient (urea)& $k_u$	&	$21.3$ & $\sfrac{\text{kg}}{\text{m}^3}$&\cite{Lauchnor:Article:2015}\\
Maximum specific growth rate &$\mu$ &    $4.17\times 10^{-5}$& $\sfrac{1}{\text{s}}$&\cite{Conolly:Article:2013}\\
Maximum rate of urea utilization &$\mu_u$ &    $1.61\times 10^{-2}$ & $\sfrac{1}{\text{s}}$&\cite{Lauchnor:Article:2015}\\
Microbial attachment rate &$k_a$ &    $8.38\times 10^{-8}$ & $\sfrac{1}{\text{s}}$&\cite{Hommel:Article:2015}\\
Microbial death rate&$k_{d}$ & 	$3.18\times 10^{-7}$ & $\sfrac{1}{\text{s}}$&\cite{Taylor:Article:1990}\\ 
Minimum permeability& $K_\text{min}$	& $10^{-20}$ & $\text{m}^2$ &\cite{Schlumberger:Definition:2020}\\
Oxygen consumption factor&$F$ & 	$0.5$ &  [-] & \cite{Mateles:Article:1971}\\ 
Water viscosity &$\mu_w$	& 	$2.54\times10^{-3}$ & $\textrm{Pa s}$ & \cite{Ebigbo:Article:2007}\\
Yield coefficient (growth) &$Y$	& 	$0.5 $ & [-] & \cite{Seto:Article:1985}\\
Yield coefficient (calcite/urea) &$Y_{uc}$	& 	$1.67 $ & [-] & Universal\\
\hline
\end{tabular}
\label{tab:allparameterss}
\end{table} 

The equivalent radius $r_e$ for the injection well depends on the grid. For a domain with rectangular grid blocks, the equivalent radius is given by $r_e=0.14\sqrt{\Delta x^2+\Delta y^2}$ \citep{Peaceman:Article:1978}. We set the well radius to r$_{\text{I}}$=0.15 m \citep{Ebigbo:Article:2007}. Regarding input concentrations, the maximum amount of urea and oxygen dissolved in water is limited by its solubility, e.g., 1079 kg/m$^3$ at 20 $\degree$C for urea and 0.04 kg/m$^3$ at 25 $\degree$C for oxygen. In the MICP experiment reported in \cite{Whiffin:Article:2007} the concentration of urea corresponds to 66 kg/m$^3$. The concentration of injected microbes is typically given in colony forming units (CFU) or in optical density of a sample at 600 nm (OD$_{600}$). Two values of concentrations for \emph{S. pasteurii} used in experiments and reported in literature are 4$\times 10^7$ CFU/ml and 1.583 OD$_{600}$. The former is equivalent to 0.01 kg/m$^3$ using a cell weight of 2.5$\times 10^{-16}$ kg/CFU \citep{Norland:Article:1987} while the latter is approximately equal to 17$\times 10^{8}$ CFU/ml \citep{Jin:Article:2018}, which, using the cell weight, is converted to 0.425 kg/m$^3$. Here we consider the following concentrations for the rate-limiting components (suspended microbes, oxygen, and urea) in the three injected solutions (microbial, growth, and cementation solutions): $c_m$=0.01 kg/m$^3$, $c_o$=0.04 kg/m$^3$, and $c_u$=300 kg/m$^3$.

Different studies can be conducted on mathematical models with a few parameters. For example, sensitivity analysis on the mathematical model allows us to identify critical model parameters. We refer to \cite{Landa:Article:2020b} for the description of a novel sensitivity analysis method. Other common studies on these models are, for example, mathematical optimization and parameter uncertainty. In \cite{Tveit:Article:2020}, we present an optimization study of a MICP model under parameter uncertainty.

\subsection{Injection strategy}\label{sec:inj_strat}
\begin{figure}[h!]
\centering
\includegraphics[width=\textwidth]{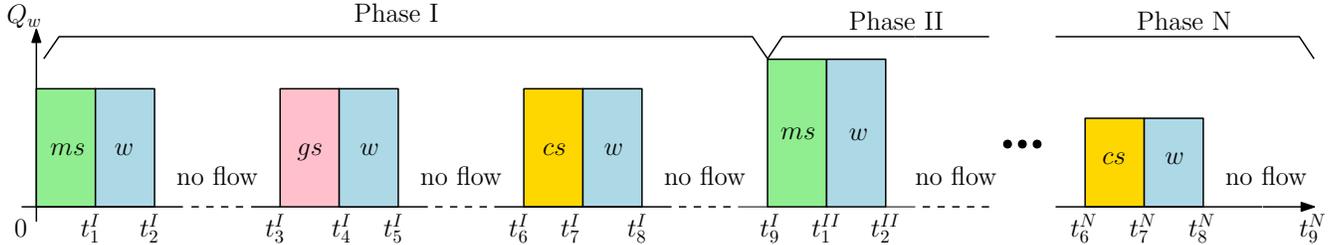}
\caption{Injection strategy splitting into phases where $ms$, $gs$, $cs$, and $w$ refer to injection of microbial, growth, and cementation solutions and only water respectively.}
\label{inj} 
\end{figure}

Diverse injection strategies have been studied for the MICP technology in laboratory experiments [e.g., \cite{Ebigbo:Article:2012}, \cite{Kirkland:Article:2019}, and \cite{Whiffin:Article:2007}] and numerical simulations [e.g., \cite{Hommel:Article:2015}, \cite{Minto:Article:2019}, and \cite{Tveit:Article:2018}]. In this work we consider the injection strategy shown in Fig. \ref{inj}.

By separating the injection of solutions (microbial, growing, and cementation solutions) with no-flow periods and considering the retention times for the different processes (bacterial attachment, biofilm formation, and calcite precipitation), limited clogging is expected to occur near the injection site \citep{Tveit:Article:2018,Whiffin:Article:2007,Yu:Article:2020}. Given that the position of the well is fixed in the domain, the control variables for the injection strategy are the flux rate (water) along the height of the well, i.e., $Q_w(z,t)$ and concentrations of the rate-limiting components (microbes, oxygen, and urea). This injection strategy involves several phases where the three solutions are injected in the following order: microbial, growth, and cementation solutions. First, microbes are injected for a total time $t_1^I$. This injection is followed by water injection ($t_2^I$) to move the suspended microbes away from the injection well. Subsequently, there is a no-flow period to facilitate attachment of suspended microbes to the pore walls ($t_3^I$). Growth solution is injected ($t_4^I$), followed by water displacement ($t_5^I$), and subsequently there is a no-flow period ($t_6^I$) to stimulate biofilm formation away from the injection well and around the sealing target. Cementation solution is injected ($t_7^I$), displaced by water ($t_8^I$), and subsequently a no-flow period ($t_9^I$) to precipitate calcite at the sealing target. We refer to these nine stages as phase I. Several phases can be applied to decrease the permeability in the target zone, see Fig. \ref{inj}.  

\section{Numerical studies}\label{numericalstudies}
In this section, we consider several examples that are divided into two parts. In the first part we study MICP in systems where we target calcite precipitation at selected parts of the aquifer [e.g., \cite{Minto:Article:2019}, \cite{Nassar:Article:2018}, and \cite{Tveit:Article:2018}]. This mimics a situation where MICP technology is applied to prevent formation of leakage paths in the caprock, that is in regions with closed fractures/faults that could be opened when CO$_2$ is injected. In the second part we study MICP in systems where leakage paths are modeled explicitly [e.g., \cite{Cunningham:Article:2019}]. Here, we focus on the benchmark problem introduced in \cite{Ebigbo:Article:2007} and \cite{Class:Article:2009}, where two aquifers are separated by a caprock with a leakage path. 

For the numerical examples we consider two set of reservoir properties. The first set of properties is taken from \cite{Tveit:Article:2018}, where the authors studied the MICP technology for sealing at a given distance of an injection well. One of the motivations to include the same reservoir properties in this work is to compare qualitatively the simulation results between the two different model implementations. The second set of properties is taken from \cite{Ebigbo:Article:2007}. Let $K_A$ denote the permeability in the aquifer, $K_L$ the permeability of the leakage path, $L$ the length, $W$ the width, and $H$ the height of the aquifer, $h$ the height of the caprock, $l$ the distance of the leakage zone from the well, $a$ the aperture of the leakage path, and $\omega$ the aperture of the potential leakage zone. Table \ref{tab:dominios} summarizes the properties of both systems.

\begin{table}[h!]
\centering
\caption{Table of reservoir properties for the numerical studies.}
\begin{tabular}{ l l l l c c c c c c c c c}		
\hline
Reference & Example & $\phi$ & $K_A$[m$^2$] & $K_L$[m$^2$] & $L$[m] & $H$[m]  & $W$[m]   & $h$[m]  & $l$[m]  & $\omega$[m] & $a$[m]\\
\hline
 &1Dfhs &  &  &  &  & & \multirow{2}{*}{---}  &  &  &  & \\
\cite{Tveit:Article:2018}&2Dfhcs & 0.2 & $10^{-12}$ & --- &75 & --- &   & --- & 10 & 5 & --- \\
\cline{8-1}
&2Dfhrs&  &  &  &  & & 20  &  &   &  & \\
\hline
&2Dfvrs&  &  &  &  & &  \multirow{2}{*}{---} & --- & 90 & 20 & ---\\
\cline{9-1}
\cline{10-1}
\cline{11-1}
\cline{12-1}
\cite{Ebigbo:Article:2007} & 2Dfls & 0.15 & $2\times10^{-14}$ & $10^{-12}$ & 500 & 30 &   & \multirow{2}{*}{100} & \multirow{2}{*}{100} & \multirow{2}{*}{---} & \multirow{2}{*}{0.3}\\
\cline{8-1}
&3Dfls& &  &  & & & 1000 &  & & & \\
\hline
\end{tabular}
\label{tab:dominios}
\end{table} 

In the examples, we will perform simulations on 1D, 2D, and 3D flow systems. On each of the systems we will study different aspects of the injection strategy (Section~\ref{sec:inj_strat}) and the dynamics of the MICP process. The learnings from one system will be useful by itself, but will also inform the studies on the other systems, ultimately leading up to running the 3D benchmark problem in the second part. Since this benchmark problem involves the solution on a large domain, we will neglect the dispersion effects to decrease the computational time (only for the 2Dfl and 3Dfl systems since we will compare their numerical results). We remark that for the numerical simulations the 1D and 2D flow systems are 3D grids (e.g., the 1D flow horizontal system is represented by a grid of dimensions \mbox{L$\times$1 m$\times$1 m).} 

\subsection{MICP to prevent formation of leakage paths} 
\begin{figure}[h!]
\centering
\includegraphics[width=\textwidth]{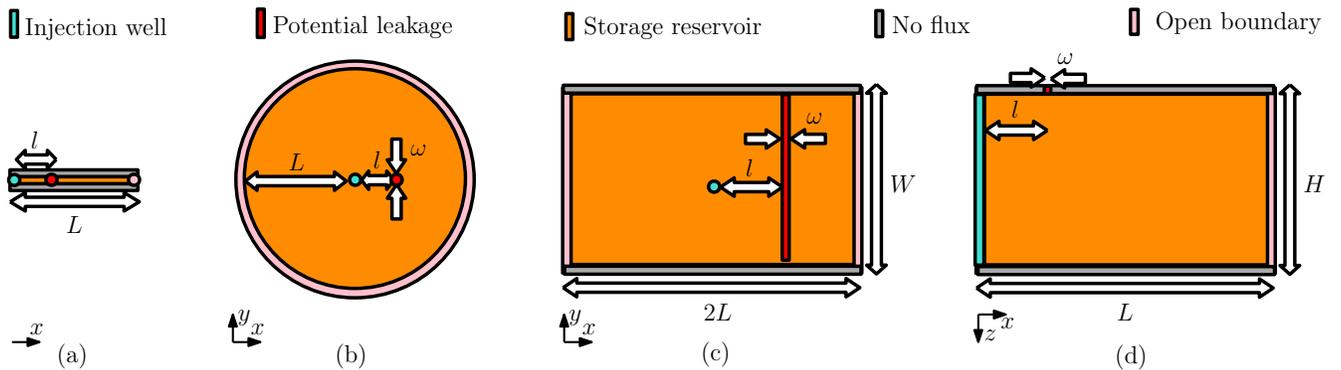}
\caption{(a) 1D flow horizontal system (1Dfhs), (b) 2D flow horizontal circular system (2Dfhcs), (c) 2D flow horizontal rectangular system (2Dfhrs), and (d) 2D flow vertical rectangular system (2Dfvrs).}
\label{horizontal2D} 
\end{figure}

Fig. \ref{horizontal2D} shows four different systems we consider for the numerical experiments. In all experiments the potential leakage region is located at a distance $l$ from the injection well. The simplest spatial domain for numerical studies is a 1D flow horizontal system as shown in Fig. \ref{horizontal2D}a. This domain consists of an injection well, a potential leakage region, and an open boundary. Two 2D flow horizontal extensions of this system are given in Fig. \ref{horizontal2D}b and Fig. \ref{horizontal2D}c. The former represents a potential leakage region with a given aperture $\omega$, while the latter represents a potential leakage region of aperture $\omega$ and width $W$. Fig. \ref{horizontal2D}d shows a 2D flow vertical system with a height $H$ where the potential leakage region is on the top caprock.

\subsubsection{1D flow horizontal system (1Dfhs)} 
\begin{figure}[h!]
\centering
\includegraphics[width=.85\textwidth]{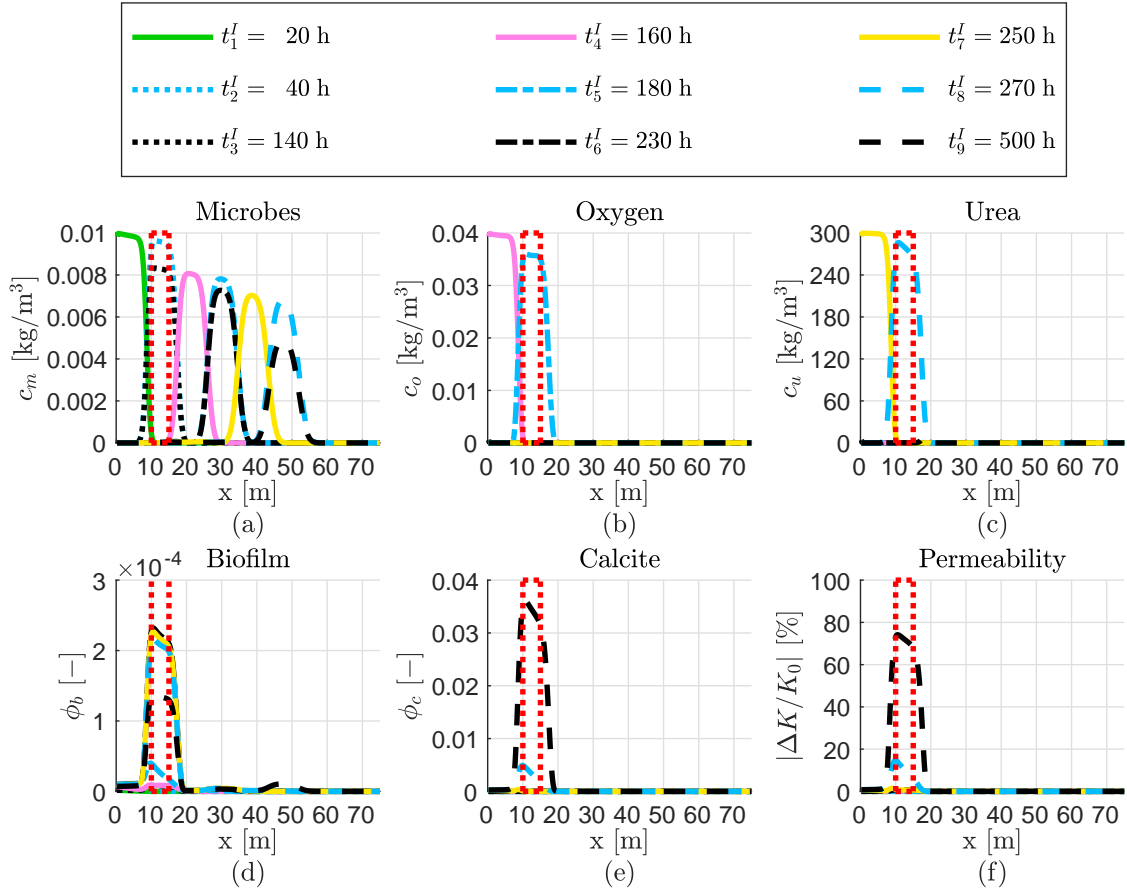}
\caption{Spatial distribution of (a) suspended microbes, (b) oxygen, (c) urea, (d) biofilm, (e) calcite, and (f) permeability at different times in the injection strategy (1Dfhs). The potential leakage region is inside the red rectangle.}
\label{micp_1DH} 
\end{figure}

We first investigate the dynamic evolution of the model components (i.e., suspended microbes, oxygen, urea, biofilm, and calcite) during the injection of phase I on the 1Dfhs in Fig. \ref{horizontal2D}a. The different values for the times in the injection of phase I are the following: $t_1^I=20$ h, $t_2^I=40$ h, $t_3^I=140$ h, $t_4^I=160$ h, $t_5^I=180$ h, $t_6^I=230$ h, $t_7^I=250$ h, and $t_8^I=270$ h. These injection times are identical to the ones studied in \cite{Tveit:Article:2018}. After performing simulations changing manually the injection rate, a value which leads to permeability reduction on the target zone is $Q^I_w$=$2.4\times 10^{-5}$ m$^3$/s. The numerical results are shown in Fig. \ref{micp_1DH}. We observe that after 500 hours all of the urea is used to produce calcite over the potential leaky zone. The pore space around the target zone is reduced significantly after injection of phase I. 

\subsubsection{2D flow horizontal circular system (2Dfhcs)} 
\begin{figure}[h!]
\centering
\includegraphics[width=.7\textwidth]{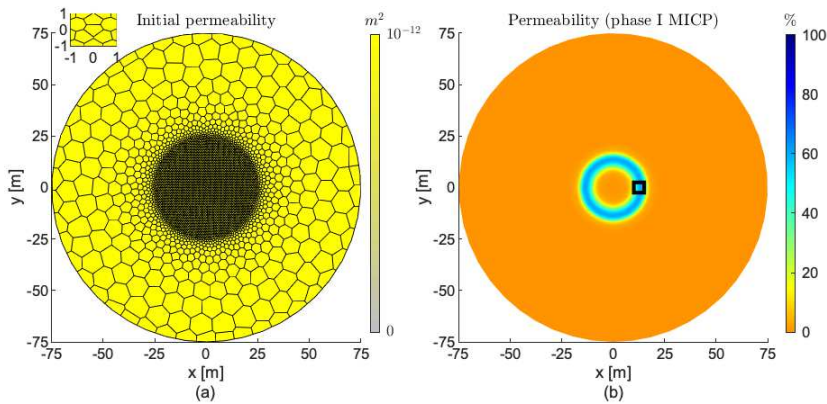}
\caption{(a) Initial permeability and (b) permeability reduction after injection of phase I (2Dfhcs). The potential leakage region is inside the black square.}
\label{comp} 
\end{figure}

We now consider the 2Dfhcs in Fig. \ref{horizontal2D}b studied in \cite{Tveit:Article:2018}. The authors used a sequential approach to solve the mathematical model on a fine triangular grid, which was implemented using FiPY \citep{Guyer:Article:2009}. The significant permeability reduction was at a distance of 10 to 15 m from the injection well, with a maximum and average permeability reductions of ca. 80\% and 60\% respectively. Most of the model parameters considered in \cite{Tveit:Article:2018} have the same values as in Table \ref{tab:allparameterss} or are of the same order of magnitude. The radius of the domain, target location of MICP, initial porosity, and permeability are the same as in the 1D experiment, which also are the mean values in the log-normal distributions in \cite{Tveit:Article:2018}. The main purpose of this example is to compare qualitatively with the results in \cite{Tveit:Article:2018}. We simulate the injection of one phase of MICP using the same injection times as in the previous example (1Dhd). Testing multiple values with simulations, an injection rate which results in reduction of permeability over the target zone is $Q^I_w$=$1.2\times 10^{-3}$ m$^3$/s. Fig. \ref{comp} shows the grid, initial permeability, and permeability reduction for our numerical simulations. 

In Fig. \ref{comp}b the significant permeability reduction is at a distance of 10 to 15 m from the injection well, with a maximum and average permeability reductions of ca. 60\% and 50\% respectively. Comparing qualitatively the permeability reduction reported in \cite{Tveit:Article:2018} to the one seen in Fig. \ref{comp}b, we observe that both simulations predict the reduction of permeability at the target distance from the injection well. We also observe that the average permeability reduction is of the same order of magnitude in both simulations. The different approaches to model some of the MICP processes [e.g., detachment from growing biofilm in \cite{Tveit:Article:2018} and detachment due to erosion in this work] results in the discrepancies between the predicted permeability reductions. In addition, the computational cost of the present grid is lower compared to the uniform fine triangular grid studied in \cite{Tveit:Article:2018}. Thus, in the subsequent experiments, we discretize the spatial domain in a similar manner as shown in Fig. \ref{comp}a, where the grid around the injection well and the region where the calcite precipitation occurs is fine (order of tens of centimeter), and gradually becomes coarser (order of meters) towards the domain boundaries. 

\subsubsection{2D flow horizontal rectangular system (2Dfhrs)} We focus on the 2Dfhrs in Fig. \ref{horizontal2D}c. We set the simulation domain size to $2L=150$ m and $W=20$ m. For this example we investigate the reduction of permeability in a potential leakage zone along the width of the aquifer. We simulate the injection of one phase of MICP using the same injection times as in the previous example. Testing multiple values with simulations, an injection rate which results in reduction of permeability over the target zone is $Q^I_w$=$7.2\times 10^{-4}$ m$^3$/s.

\begin{figure}[h!]
\centering
\includegraphics[width=\textwidth]{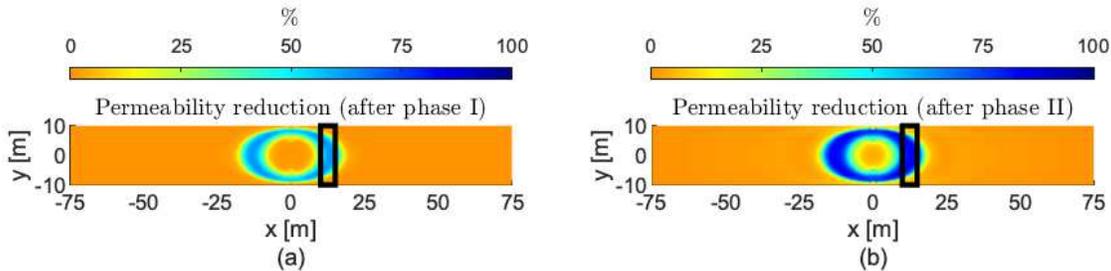}
\caption{Permeability reduction after injection of (a) phase I and (b) phase II (2Dfhrs). The potential leakage region is inside the black rectangle.}
\label{2DH_rectangle} 
\end{figure}

Fig. \ref{2DH_rectangle}a shows the permeability reduction after phase I of the injection. We observe that the closer to the lateral boundaries we target the calcite precipitation, the further into the aquifer the components need to be injected, due to the radial flow. Consequently, not all parts of the potential leakage region are covered by one phase of MICP injection. We apply a second phase of injection with the same injection rate, concentrations, and time intervals as phase I; see Fig. \ref{2DH_rectangle}b. We observe that after phase II the reduction of permeability is greater; however, the areas close to the boundaries inside the potential leakage region are not reached. Thus, several injection phases at different rates are needed to reduce the permeability inside the potential leakage region. 

\subsubsection{2D flow vertical rectangular system (2Dfvrs)} In the next example we study the 2Dfvrs shown in Fig. \ref{horizontal2D}d. We set the simulation domain size following the benchmark study in \cite{Ebigbo:Article:2007}, that is, $L=500$ m and $H=30$ m. We investigate two different injection approaches along the well for the components and water to efficiently get calcite precipitation at the potential leakage region in the caprock. For the first simulation the injection of components and water is only in the first 3 m of the well (strategy A). For the second simulation we change the water injection to be along the whole height of the well, but all MICP components are still only injected at the top of the well (strategy B). Given that the distance to the leakage zone for this reservoir is larger than the one in the previous examples, we change the injection rate and times. After performing simulations changing manually these values, the following values lead to permeability reduction over the target zone: $t_1^I=15$ h, $t_2^I=26$ h, $t_3^I=100$ h, $t_4^I=130$ h, $t_5^I=135$ h, $t_6^I=160$ h, $t_7^I=200$ h, $t_8^I=210$ h, and $Q^I_w=5\times10^{-3}$ m$^3$/s. 

\begin{figure}[h!]
\centering
\includegraphics[width=\textwidth]{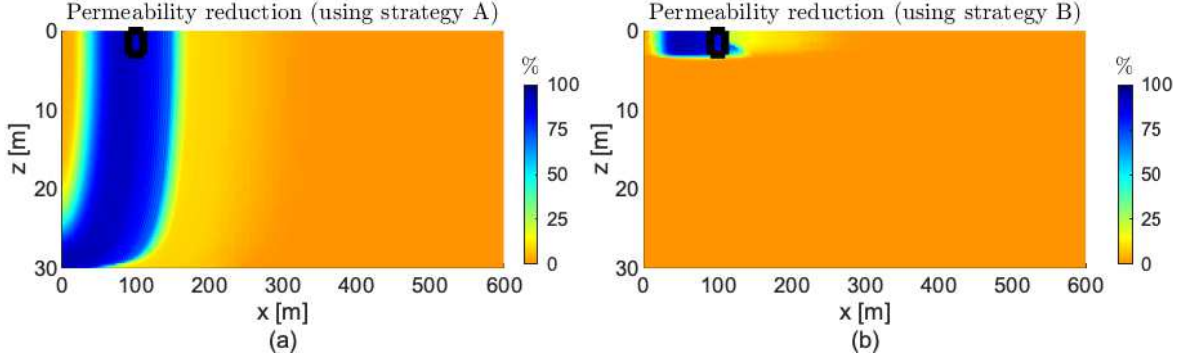}
\caption{Permeability reduction (a) only using the top of the well and (b) the whole well (2Dfvrs). The potential leakage region is inside the black rectangle.}
\label{permver2D} 
\end{figure}

Fig. \ref{permver2D} shows the permeability reduction for both injection approaches. We observe that strategy A results in calcite precipitation also along the vertical direction. This is not desired as it could lead to encapsulation of the injection well. With strategy B, we accomplish calcite precipitation only around the potential leakage region located near the caprock. Then we consider strategy B in the next examples where vertical wells are also simulated. The difference between the predicted permeability reduction in Figs. \ref{permver2D}a and \ref{permver2D}b is due to the different flow fields. In Fig. \ref{permver2D}a the injection is only at the top of the well, leading to the injected components being spread over the whole height of the reservoir. In Fig. \ref{permver2D}b the water injection through the whole height of the well keeps the flow field horizontal, forcing the injected components to flow at the top of the reservoir. We recall that this model assumes a constant-composition independent density, i.e., the density of water does not depend on the component concentrations.  

\subsection{MICP to seal leakage paths}
\begin{figure}[h!]
\centering
\includegraphics[width=.9\textwidth]{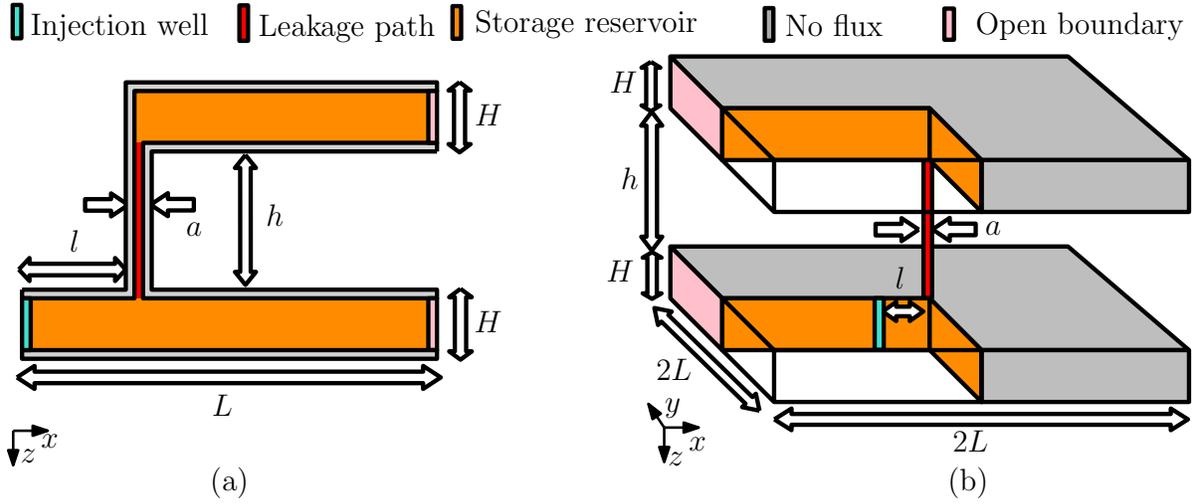}
\caption{(a) 2D flow leaky system (2Dfls) and (b) 3D flow leaky system (3Dfls).}
\label{all3D} 
\end{figure}

Diverse reservoir representations where leakage paths are explicitly modeled can be found in literature. In this work, we focus on the two domains shown in Fig. \ref{all3D}. A simple representation of a 2D flow system with one leakage path between two aquifers is shown in Fig. \ref{all3D}a. A well-established 3D benchmark for CO$_2$ leakage is given by the domain in Fig. \ref{all3D}b \citep{Class:Article:2009,Ebigbo:Article:2007}.

\subsubsection{2D flow leaky system (2Dfls)} We focus on the domain shown in Fig. \ref{all3D}a. In \cite{Ebigbo:Article:2007} the leakage is given as a result of a well which is modeled as a porous medium with higher permeability than the formation. To asses the leakage rate of CO$_2$ before and after application of MICP, we solve a simple two-phase flow model for CO$_2$ and water (see \ref{appendixA}).

\begin{figure}[h!]
\centering
\includegraphics[width=\textwidth]{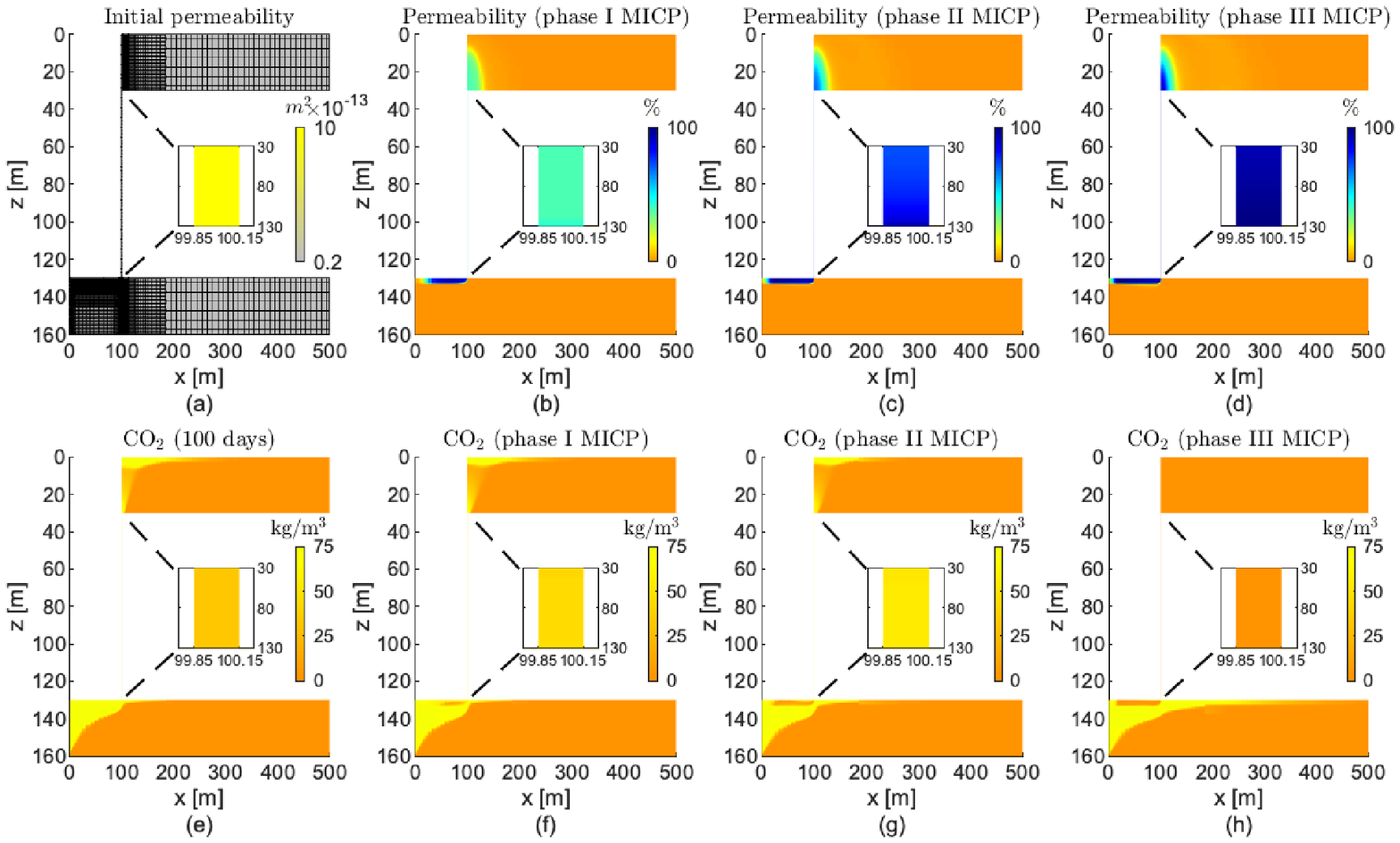}
\caption{(a) Initial permeability, (b-d) permeability reductions after phase I, II and III, and (e-h) amount of CO$_2$ in the four different scenarios after 100 days of injection (2Dfls).}
\label{meetings} 
\end{figure}

Since performing simulations on this 2D flow system is computationally cheap, we proceed to design an injection strategy for the sealing of the leakage path. It is beyond the scope of this paper to perform optimization studies. Here we use an ad-hoc approach where we keep the same values of concentrations, injection rate, and height of injection along the well as in the previous example (2Dfvrs). We set all values of time in phase I as in the previous example (2Dfvrs). Using insight gained from previous studies, we perform several simulations where injection times for the subsequent phases are changed manually. The following times lead to the sealing of the leakage path after injection of three phases: $t_4^{II}=630$ h, $t_5^{II}=650$ h, $t_6^{II}=670$ h, $t_7^{II}=690$ h, $t_8^{II}=710$ h, $t_9^{II}=800$ h, $t_7^{III}=820$ h, $t_8^{III}=840$ h, and $t_9^{III}=950$ h. Note that in this strategy microbes are not injected in phases II and III and there is only injection of urea in phase III. Fig. \ref{meetings} shows the numerical results of this injection strategy on the 2Dfls.

\begin{figure}[h!]
\centering
\includegraphics[width=.5\textwidth]{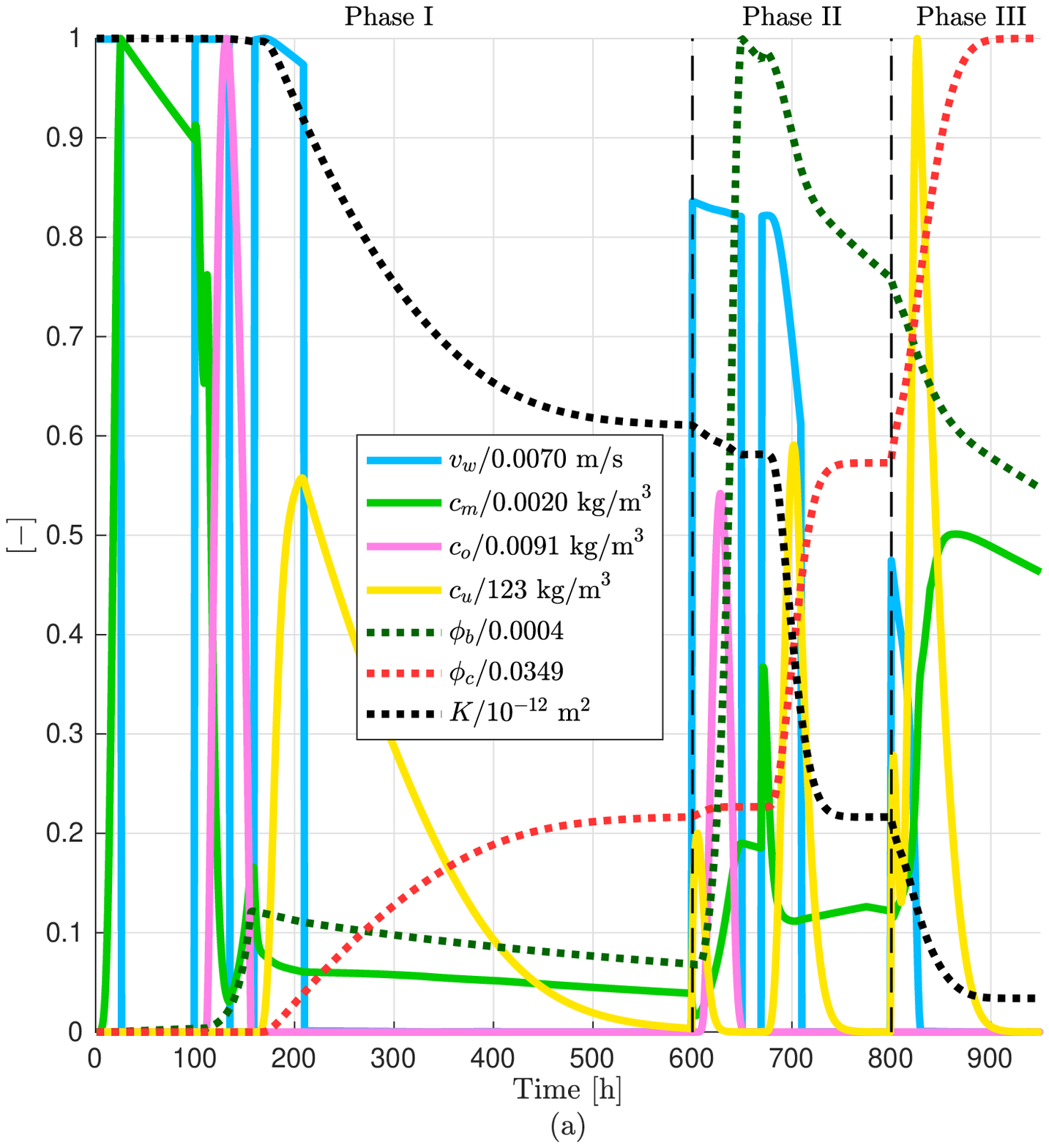}\includegraphics[width=.5\textwidth]{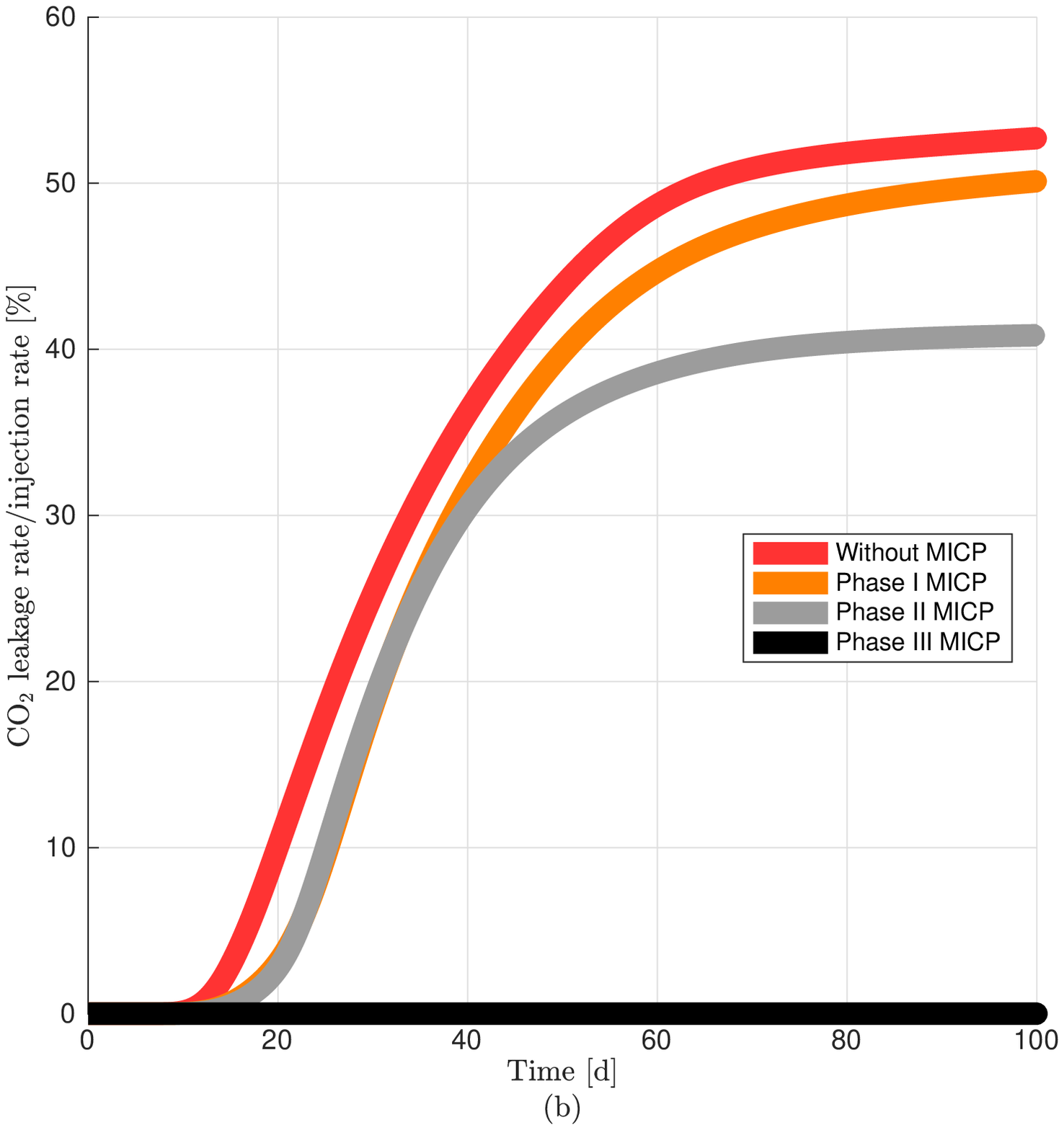}
\caption{(a) Normalized average variables (along the leakage path) and (b) leakage rate of CO$_2$ through the leakage path (at $z$ = 80 m) in the caprock (2Dfls).}
\label{co2300} 
\end{figure}

For a better visualization of the different MICP processes, in Fig. \ref{co2300}a we plot the average value normalized by its maximum value achieve in phase I, II, or III for the discharge per unit area, microbial, oxygen, and urea concentrations, biofilm and calcite volume fractions, and permeability reduction in the leakage path. We observe a remarkable increase of calcite after injection of urea in phase III which in turn decreases significantly the volume fraction of biofilm. Fig. \ref{co2300}b shows the leakage rate without and after MICP injection of phase I, II, and III after 100 days of CO$_2$ injection. In the numerical results, we calculate the leakage as the CO$_2$ flux at the middle of the leaky well ($z$ = 80 m) \citep{Ebigbo:Article:2007}. We observe that the leakage rate is practically zero after three phases.

\subsubsection{3D flow leaky system (3Dfls)} 
\begin{figure}[h!]
\centering
\includegraphics[width=\textwidth]{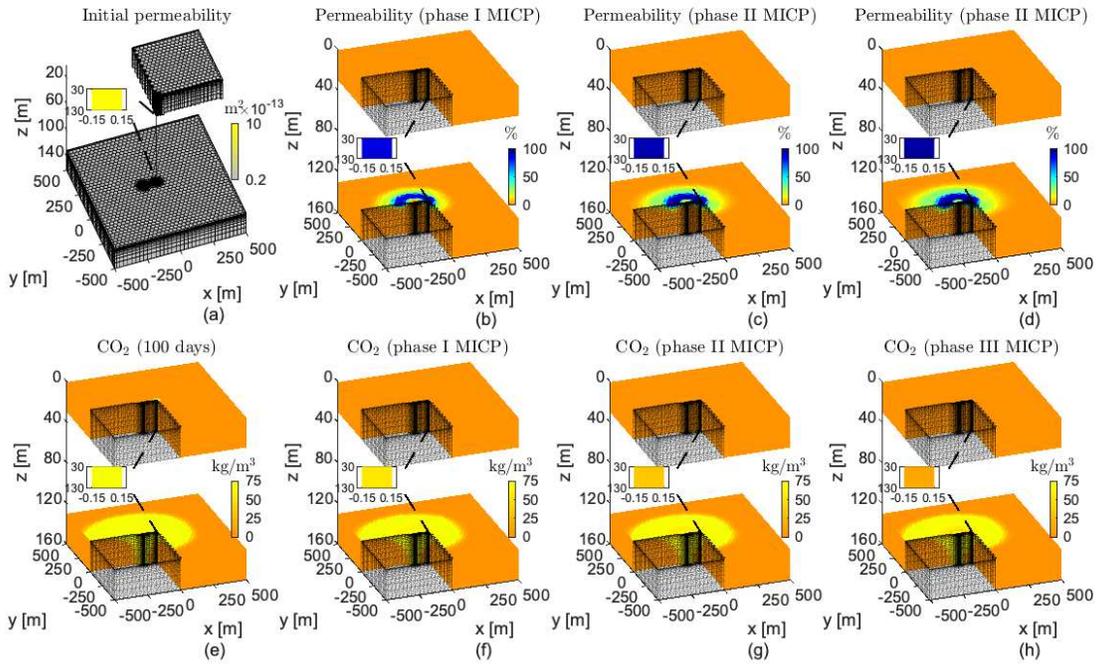}
\caption{(a) Initial permeability, (b-d) permeability reductions after phase I, II, and III, and (e-h) amount of CO$_2$ in the four different scenarios after 100 days of injection (3Dfls).}
\label{3dbenchmicp} 
\end{figure}

We consider the 3D benchmark reservoir as described in \cite{Class:Article:2009} and \cite{Ebigbo:Article:2007} shown in Fig. \ref{all3D}b. Since the properties of the previous example (2Dfls) are also equal to the ones in the benchmark, we expect to obtain similar results after applying the same injection strategy. Thus, we simulate the injection of three phases of MICP using identical time intervals as in the previous example. We set the injection rate equal to $Q^I_w$=3 m$^3$/s. Fig. \ref{3dbenchmicp} shows the numerical results after applying phase I, II, and III of MICP. 

\begin{figure}[h!]
\centering
\includegraphics[width=.5\textwidth]{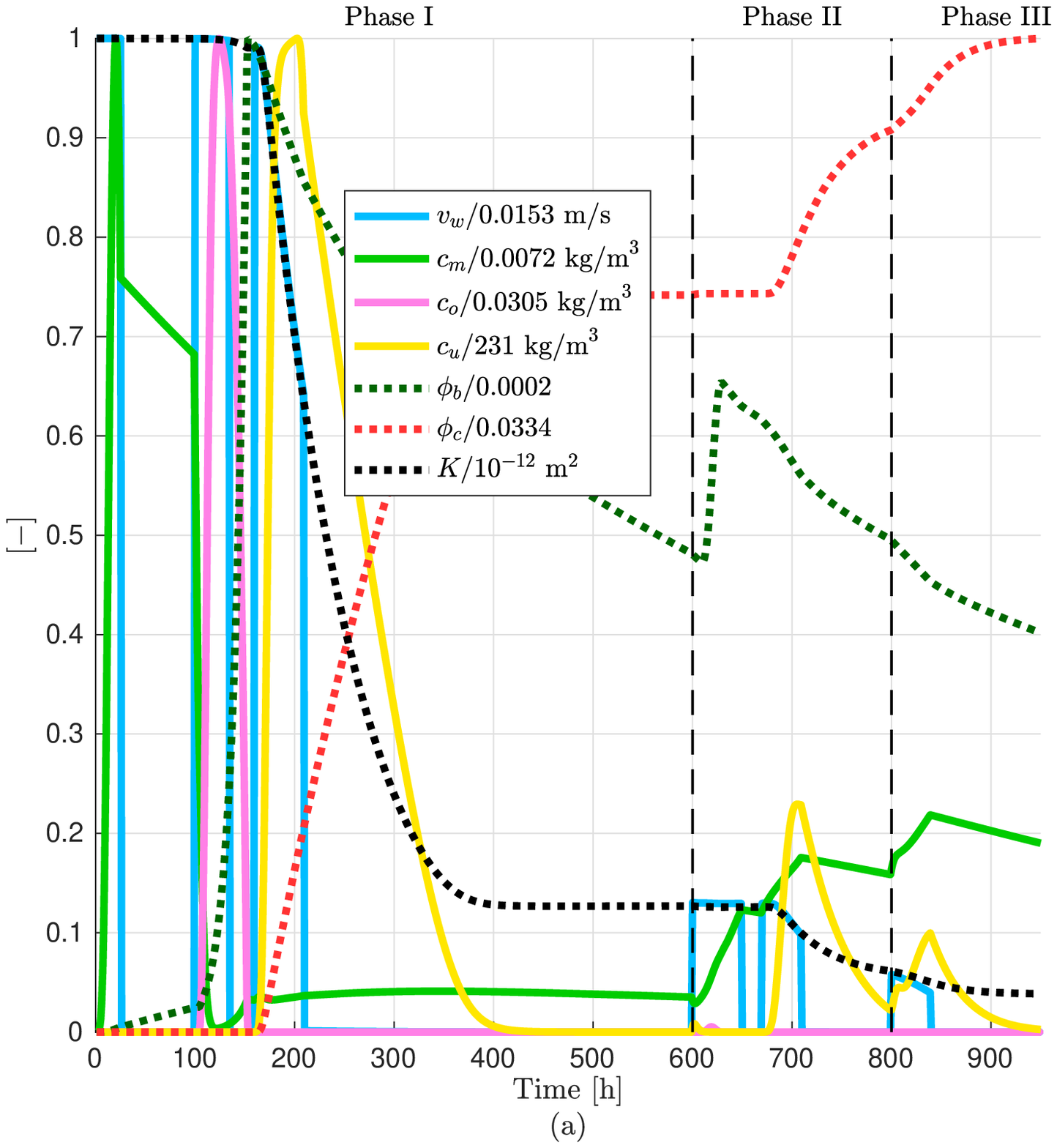}\includegraphics[width=.5\textwidth]{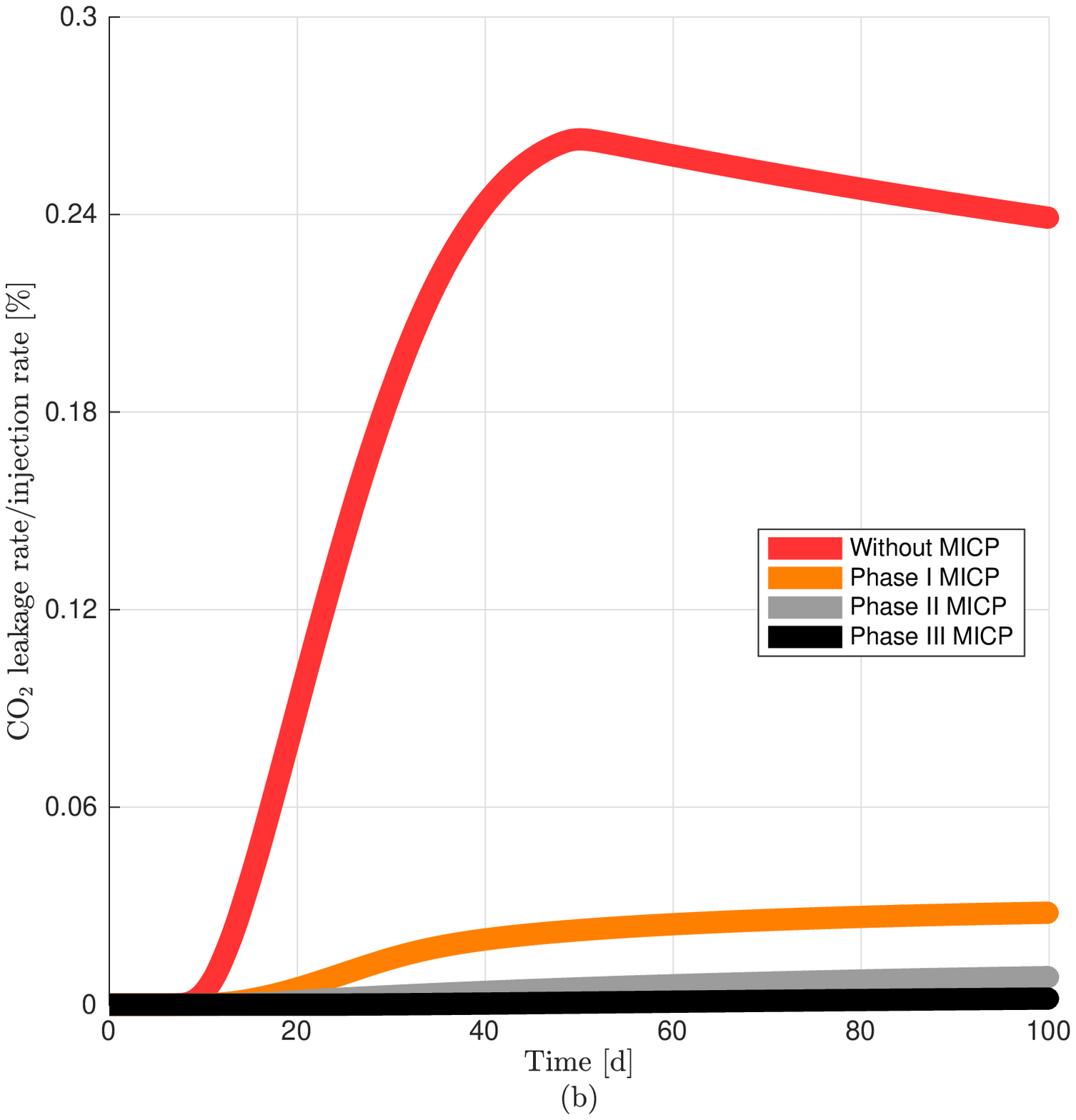}
\caption{(a) Normalized average variables (along the leakage path) and (b) leakage rate of CO$_2$ through the leakage path (at $z$ = 80 m) in the caprock (3Dfls).}
\label{benchebibo} 
\end{figure}

Fig. \ref{benchebibo} shows the different MICP processes at the leaky well and the leakage rate before and after MICP treatments. We observe that the dynamics of the processes are similar to the ones plotted for the 2Dfls. We also observe that the curve without MICP injection is in good agreement with the ones presented in the benchmark study for CO$_2$ leakage in \cite{Class:Article:2009}. Then, as observed in the 2Dfls, the leakage stops after applying three phases of MICP treatment.

\section{Discussion}
In Table \ref{tab:sum} we sum up the results in Section \ref{numericalstudies}.

\begin{table}[h!]
\centering
\caption{Injection rates, times, and permeability reductions of the numerical examples.}
\begin{tabular}{ l l l l l l l}		
\hline
Reservoir properties & Example & Rate [m$^3$/s] & Phase I [h] & Phase II [h] & Phase III [h] & 1-K/K$_0$ [\%]\\
\hline
 &  &  & $t_1^I\;\;\;$ $t_2^I\;\;\;$ $t_3^I$ & $t_1^{II}\;$ $t_2^{II}\;$ $t_3^{II}$ & $t_1^{III}$ $t_2^{III}$ $t_3^{III}$ & \\
Reference & Name & $Q_w$ & $t_4^I\;\;\;$ $t_5^I\;\;\;$ $t_6^I$ & $t_4^{II}\;$ $t_5^{II}\;$ $t_6^{II}$ & $t_4^{III}$ $t_5^{III}$ $t_6^{III}$ & mean/max \\
 &  &  & $t_7^I\;\;\;$ $t_8^I\;\;\;$ $t_9^I$ & $t_7^{II}\;$ $t_8^{II}\;$ $t_9^{II}$ & $t_7^{III}$ $t_8^{III}$ $t_9^{III}$ & \\
\hline
& &  &  &  \multirow{6}{*}{\qquad --- }  & &  \\
&1Dfhs & 2.4$\times 10^{-5}$ & & & \qquad  & 72/74 \\
& &  &  &  & &  \\
& &  & 20\;\; 40\; 140 &  & &  \\
\cite{Tveit:Article:2018}&2Dfhcs & 1.2$\times 10^{-3}$ & 160 180 230 &  & \qquad --- & 53/61 \\
& &  & 250 270 500 &  & &  \\
\cline{5-1}
& &  &  & 520 540 640  & &  \\
&2Dfhrs & 7.2$\times 10^{-4}$ &  & 660 680 730 & \qquad & 54/85 \\
& &  &  & 750 770 1000 & &  \\
\hline
& & \multirow{6}{*}{5$\times 10^{-3}$} &  &  & &  \\
&2Dfvrs &  &  & \qquad ---& \qquad ---& 82/90 \\
& &  &  & & &  \\
\cline{5-1}
\cline{6-1}
& &  & 15\;\; 26\; 100 &   &&  \\
\cite{Ebigbo:Article:2007}&2Dfls &  & 130 135 160 & \multirow{2}{*}{---\;\; ---\;\; ---} & \multirow{2}{*}{---\;\; ---\;\; ---}  & 97/$\approx$100 \\
& &  & 200 210 600 & \multirow{2}{*}{630 650 670} & \multirow{2}{*}{---\;\; ---\;\; ---}  &  \\
\cline{3-1}
& &  &  & \multirow{2}{*}{690 710 800} & \multirow{2}{*}{820 840 950} &  \\
&3Dfls & 3 &  &  & & 96/$\approx$100 \\
& &  &  & &  &  \\
\hline
\end{tabular}
\label{tab:sum}
\end{table} 

The first part of the numerical examples includes MICP simulations to prevent the formation of leakage paths in an aquifer. We first study the spatial distribution of the diverse MICP variables (namely suspended microbes, oxygen, urea, biofilm, and calcite) on a simple 1D flow system. Simulations on this system are suitable for testing large numbers of injection strategies as it requires the lowest running time. In the 2Dfhcs the potential leaky zone is given by a small area at a given distance from the injection well. However, the fluid injection through the well is in all horizontal directions leading to calcite precipitation around the injection well. To only target the potential leaky zone then the direction of injection could be controlled to decrease the cost of injected components and to not reduce the aquifer storage capability in regions unnecessary to seal. Whereas this is an evident observation, we could not find experimental nor numerical studies involving directional wells in the literature. Thus, including directional wells in the simulators could be useful for MICP studies. Since the flow velocities near the injection side are higher in radial flow (e.g., 2Dfhcs) than in plug flow (e.g., 1Dfhs), then in the former the Damköler numbers are lower in this region which effects the MICP process [see e.g., \cite{Zambare:Article:2019}]. This can be observed in Figs. \ref{micp_1DH}f and \ref{comp}b, where in the former the mean permeability reduction is higher (ca. 70\%) in comparison to the latter (ca. 60\%). An important observation of the simulations on the 2Dfhrs is that several phases of injections might be needed to precipitate calcite along the width of the aquifer as a result of the radial flow from the well. The last example in the first part is the 2Dfvrs. Here we study the injection of water and components along the well. When we inject only at the top of the well, the calcite ``encapsulates" the injection well along the vertical direction. We can mitigate this by injecting the components only at the top and water along entire height of the well which results in calcite precipitation only around the caprock. In the numerical studies we designate an arbitrary fixed part of the well for the injection of the components. However, the choice of this control variable is likely to have a significant impact in the simulations (e.g., due to transversal dispersion of the components). Hence, diverse injection heights should be studied for different systems to cut the injection times/cost.

The second part of the numerical examples focus on MICP simulations on domains where the leakage paths are modeled explicitly. We first study the sealing of a leakage path in a caprock between two aquifers on a simple 2Dfls. We proceed to find values of time intervals and injection rates to achieve sealing in the leakage path. The designed injection strategy leads to sealing of the leakage path after three phases of MICP injection. The last numerical experiment is performed on a 3D benchmark leaky well. Since reservoir properties of both domains are  equal, we use the same injection strategy as in the 2Dfls. At the end of the simulations, we observe that the leakage path is blocked successfully. Despite the satisfactory and straightforward application of the injection strategy from the 2Dfls to the 3Dfls, this is ultimately restricted by the simplicity of the system. For instance, a different injection strategy should be designed for a 3D problem with a fracture across the width of the caprock. As observed in the simulations on the 2Dfhrs in Fig. \ref{2DH_rectangle}, we expect that several phases of MICP injection are required to significantly reduce the permeability in the leakage path.

The MICP model presented in this paper is built from a simplified description of the underlying processes based on previous publications as described in Section~\ref{sec:MICP}. The mathematical model involves few equations (six mass balance equations and six cross coupling constitutive relationships) and input parameters (twenty-two parameters). Though beyond the scope of this paper, comparing numerical simulations to laboratory experiments is needed to evaluate the predictive capabilities of this simplified MICP model. Extending the model to two-phase flow (water+MICP and CO$_2$) lets us consider additional processes, e.g., dissolution of calcite due to the presence of CO$_2$; however, solving simultaneously all equations would limit the size of the problem as discussed in \cite{Cunningham:Article:2019}. As proposed by the authors, this could be solved by applying multi-physics methods, e.g., using analytical solutions for the flow.   

Differences between the leakage rate curves in Figs. \ref{co2300}b and \ref{benchebibo}b arise from different issues. Prior to MICP treatment, the percentage of leakage rate on the 2Dfls is nearly 50\% while for the 3Dfls is lower than 0.25\%. The reason for this difference is that on the 2Dfls all CO$_2$ injected either flows under or through the leakage path while for the 3Dfls only a small portion of CO$_2$ flows under and through the leakage path (since the flow is radial in the 3D system). Nevertheless, we would expect to have a similar curve shape for both systems. For the 3Dfls we observe a sharp rise of CO$_2$ leakage when the CO$_2$ reaches the leakage zone (approximately after 10 days of injection) and then drops (approximately after 50 days of injection). This is attributed to boundary effects \citep{Ebigbo:Article:2007}, as the CO$_2$ leakage is expected to continue increasing slowly after the initial sharp rise as shown in Fig. \ref{co2300}b \citep{Nordbotten:Article:2005}.

We observe a different decrease of CO$_2$ leakage after one and two phases of MICP treatment in both systems. While the permeability reduction after two phases for the 2Dfls is ca. 20\% (Fig. \ref{co2300}b), for the 3Dfls system is greater than 90\% (Fig. \ref{benchebibo}b). In this work we describe the 2Dfls as a general simple system to study MICP in a leakage path. This domain is commonly related to an approximation of a system with a fault across the caprock [e.g., see \cite{Tavassoli:Article:2018}]. In addition, the velocity close to the well in the 3Dfls is much higher than for the 2Dfls since the former is a radial flow while the latter is a linear flow \citep{Zambare:Article:2019}. Thus, numerical simulations in both systems (2Dfls and 3Dfls) lead to different flux rates in the leakage path which in turn result in different permeability reduction after application of phase I (this explains the differences between discharges per unit area, concentrations, volume fractions, and permeability reduction values in Figs. \ref{co2300}a and \ref{benchebibo}a). For these examples we observe that $v_w$ through the leakage path is slower in the 2Dfls than in the 3Dfls.  As a consequence of the difference between discharges per unit area, the whole MICP process is slowed down in the 2Dfls in comparison to the 3Dfls. In addition, from Fig. \ref{meetings} we observe calcite precipitation between the injection and leaky well while from Fig. \ref{3dbenchmicp} we observe calcite precipitation outside this region, i.e., between the injection well and outer boundary, since the flow field in the 3Dfls also transports the different components around the leakage zone. Then the difference between both flow fields have an impact on the MICP processes in the leakage path as observed in Figs. \ref{co2300}a and \ref{benchebibo}a. 

Notwithstanding these differences, the leakage path is successfully remediated in both systems (2Dfls and 3Dfls) after application of the third MICP treatment. A comprehensive investigation of boundary and grid effects, in addition to studies of more complex 3D problems, is not feasible using the current implementation of the mathematical model. Here we have used MRST for the testing of the model as implementation on this framework is not difficult; however, there are computation time limitations using this toolbox. Our current plan is to implement this mathematical model using the open porous media (OPM) initiative which allows to perform more computationally challenging simulations \citep{Rasmussen:Article:2020}. Hence we will use the OPM simulator to study these effects and MICP in more complex 3D systems.         

In this work we only study one injection strategy for the MICP application in each of the flow systems consisting of several periodical phases. As described above, each phase involves the injection of three solutions, injection of only water, and periods of no flow, given a total of 18 control variables: three concentrations, six injection rates, and nine period times. To enable other researchers to benefit from our work and test different injection strategies, we have made the code available through a repository. The links to download the corresponding code can be found above the references at the end of the manuscript. We observed that after injection of phase I, consisting of injecting the microbial, growth, and cementation solutions, separated by no-flow periods (Fig. \ref{inj}), avoids major calcite precipitation near the injection site (see e.g., Fig. \ref{micp_1DH}e). This was not the case when we sealed the leakage path where additional injection of phases was required (see e.g., Fig. \ref{meetings}d). 

The modeling work in this study highlights the importance of numerical modeling for a successful implementation of MICP in practice. The workflow presented here should be combined with optimization to design effective field strategies for plugging leakage pathways at a considerable distance from the injection well. For instance, optimization could be applied to maximize calcite precipitation in the leakage path while minimizing costs and the calcite precipitation between the injection well and leakage path. After injection of phase I, changing the sequence of components for the subsequent phases could lead to a better injection strategy, e.g., injection of growth solution after the first phase for microbial resuscitation or only injection of cementation solution if there is still enough biofilm left. Then, studying different sequence of components plus different heights of the well adds more complexity to the optimization. Though beyond the scope of the current work, further study in this direction is needed for optimization of the MICP technology.

\section{Conclusions}
In this work, we present a simplified mathematical model for the MICP technology including the transport of injected solutions (microbial, growth, and cementation solutions), biofilm formation, and calcite precipitation. This model is developed for computational efficiency, to accommodate for large computational domains and optimization problems in field-scale simulations. We study an injection strategy involving several phases where each phase includes microbial, growth, and cementation solutions with periods of only injection of water and no flow. We conduct diverse numerical experiments for various one- and two-dimensional flow systems to study the MICP process. Finally, we show the results of applying MICP to a well-established 3D benchmark problem for CO$_2$ leakage.  

Based on this work our conclusions are as follows:
\begin{itemize}
\item Learnings from 1D and 2D flow studies help us to develop practical injection approaches for 3D simulations. 
\item Only using a part of the well for injection of components and water leads to calcite precipitation along the whole vertical direction; however, using the top part of the well for injection of components and the rest of the well for water injection leads to calcite precipitation on the top of the aquifer.    
\item Several phases of injection might be needed for decreasing the permeability in (potential) leakage regions as a result of the radial flow by the injection well.
\item This study demonstrates that it is possible to use MICP technology to plug a leakage pathway even at considerable distance from the injection well (order of 10 to 100 meters) if the location of the leakage pathway is well known.
\end{itemize}

\setcounter{equation}{0}
\setcounter{table}{0}
\setcounter{section}{0}
\renewcommand{\theequation}{A-\arabic{equation}}
\renewcommand{\thetable}{A-\arabic{table}}
\renewcommand\thesection{Appendix \Alph{section}}
\section{Two-phase flow mathematical model for CO$_2$ and water}\label{appendixA}
We describe the simplified two-phase flow model used in \cite{Class:Article:2009} for a benchmark study in problems related to CO$_2$ storage. CO$_2$ and water are assumed immiscible and incompressible. We denote by $s_w$ the water saturation and $s_\text{CO$_2$}$ the CO$_2$ saturation ($s_w+s_\text{CO$_2$}=1$). We write Darcy's law and the mass conservation equations for each $\alpha$ phase ($\alpha=w,\;\text{CO$_2$}$)
\begin{equation}
\pmb{v}_{\alpha}=-\frac{k_{r,\alpha}}{\mu_\alpha}\mathbb{K}(\nabla p_\alpha-\rho_\alpha\pmb{g}),\quad \phi\frac{\partial s_\alpha}{\partial t}+\nabla\cdot\pmb{v}_\alpha=q_\alpha
\end{equation}
where $k_{r,\alpha}$ is relative permeability. The relative permeabilities are set as a linear function of the saturations ($k_{r,\alpha}=s_\alpha$), and capillary pressure is neglected ($p_w=p_\text{CO$_2$}$). The model parameters are summarized in Table \ref{co2}.
\begin{table}[h!]
\centering
\caption{Table of input variables and model parameters for the CO$_2$ leakage assessment.}
\begin{tabular}{ l l l l}		
\hline
Parameter & Symbol & Value & Unit\\
\hline
CO$_2$ viscosity &$\mu_\text{CO$_2$}$	& 	$3.95\times10^{-5}$ & $\textrm{Pa s}$\\
CO$_2$ density &$\rho_\text{CO$_2$}$	& 	$479$ & $\sfrac{\text{kg}}{\text{m}^3}$\\
Injected CO$_2$ & $Q_\text{CO$_2$}$ & 1.6 $\times10^{3}$& $\sfrac{\text{m}^3}{\text{day}}$\\
Total time of injection & T$_f$ & 100 & days\\
\hline
\end{tabular}
\label{co2}
\end{table}

\section*{Notation}
\vspace{-.5cm}
\begin{longtable}{l l}
$a$ & aperture of the leakage path, [m]\\
$c_m$, $c_o$, $c_u$ & suspended microbial, oxygen, and urea concentrations, [kg/m$^3$]\\
$\mathbb{D}_m$, $\mathbb{D}_o$, $\mathbb{D}_u$& suspended microbial, oxygen, and urea dispersion coefficients, [m$^2$/s]\\
$D_m$, $D_o$, $D_u$& suspended microbial, oxygen, and urea diffusion coefficients, [m$^2$/s]\\
$F$ & oxygen consumption factor\\
$\pmb{g}$ & gravity, [m/s$^2$]\\
$H$, $h$ & heights of the aquifer and caprock, [m]\\
$\pmb{J}_m$, $\pmb{J}_s$, $\pmb{J}_u$ & suspended microbial, oxygen, and urea fluxes, [kg/(s m$^2$)]\\
$\mathbb{K}$, $K$ & rock permeability (tensor and scalar), [m$^2$]\\
$K_A$, $K_L$, $K_\text{min}$ & aquifer, leakage, and minimum permeabilities, [m$^2$]\\
$k_a$, $k_d$ & suspended microbial attachment and death rates, [1/s]\\ 
$k_{r,\text{CO$_2$}}$,  k$_{r,w}$ & relative permeabilities of CO$_2$ and water\\
$k_o$, $k_u$ & half-velocity coefficients (oxygen and urea), [kg/m$^3$]\\
$k_{str}$ & detachment rate, [m/(s Pa)]\\
$k_{ub}$ & mass ratio of urease to biofilm\\
$k_{\text{urease}}$ & maximum activity of urease, [1/s]\\
$L$, $l$ & size of the domain and distance from the well to the leakage region, [m]\\
$p_\text{CO$_2$}$, $p_w$ & CO$_2$ and water pressures, [Pa]\\
$p_I$ & pressure inside the wellbore, [Pa]\\
$Q_{CO_2}$, $Q_w$ & injection rates of CO$_2$ and water, [m$^3$/s]\\
$q_{CO_2}$, $q_w$ & source/sink terms of CO$_2$ and water, [1/s]\\
$R_m$, $R_o$, $R_u$ & suspended microbial, oxygen, and urea rates, [kg/(s m$^3$)]\\
$r_I$ & radius of the well, [m]\\
$s_\text{CO$_2$}$, s$_w$ & CO$_2$ and water saturation\\
$t^N_{n}$ & injection time n of phase N, [s]\\ 
$T_f$ & total time of CO$_2$ injection, [s]\\ 
$\pmb{v}_\text{CO$_2$}$, $\pmb{v}_w$ & CO$_2$ and water discharges per unit area, [m/s]\\
$\pmb{v}$ & effective velocity of water, [m/s]\\
$Y$, $Y_{uc}$ & yield coefficients (growth and urea to calcite)\\
$z_{bh}$ & reference depth, [m]\\
$\alpha_{L}$, $\alpha_{T}$ & longitudinal and transverse dispersion coefficients, [m] \\
$\eta$ & fitting factor (permeability-porosity relationship)\\
$\mu_\text{CO$_2$}$, $\mu_w$ & CO$_2$ and water viscosities, [Pa s]\\
$\mu$ & maximum specific growth rate, [1/s]\\
$\mu_u$ & maximum rate of urea utilization, [1/s]\\
$\omega$ & aperture of the potential leakage zone, [m]\\
$\phi$ & rock porosity\\
$\phi_b$, $\phi_c$ & volume fractions of biofilm and calcite\\
$\phi_\text{crit}$ & critical porosity\\
$\Xi$ & length of the grid block in the major direction of the wellbore, [m]\\
$\rho_b$, $\rho_c$, $\rho_\text{CO$_2$}$, $\rho_w$ & biofilm, calcite, CO$_2$, and water densities, [kg/m$^3$]\\
\end{longtable}

\paragraph{Acknowledgements:} This work was supported by the Research Council of Norway [grant number 268390]. 

\paragraph{Data availability:} The numerical simulator MRST used in this study can be obtained at \href{https://www.sintef.no/projectweb/mrst/}{https://www.sintef.} \href{https://www.sintef.no/projectweb/mrst/}{no/projectweb/mrst/}. Link to complete codes for the numerical studies can be found in \href{https://github.com/daavid00/ad-micp.git}{https://github.com/} \href{https://github.com/daavid00/ad-micp.git}{daavid00/ad-micp.git}. The mesh generator DistMesh used for the spatial discretization in some of the numerical studies can be found in \href{http://persson.berkeley.edu/distmesh/}{http://persson.berkeley.edu/distmesh/}.

\end{document}